\documentclass[a4paper,11pt]{article}
\pdfoutput=1 

\usepackage{jheppub} 
\usepackage{slashed}
\usepackage[normalem]{ulem}
\usepackage{braket}
\usepackage[T1]{fontenc} 

\title{\boldmath Krylov complexity and spectral density of BMN matrix model}

\author[a]{Dibakar Roychowdhury}

\affiliation[a]{Department of Physics, Indian Institute of Technology Roorkee,\\Roorkee 247667, Uttarakhand, India}

\abstract{We use Krylov complexity as a typical diagnostic of quantum many body dynamics in the context of BMN matrix model at large mass gap. We calculate several physical entities, for example, the moments and return amplitudes at large mass deformation. We calculate them for both spread complexity of states as well as Krylov growth of operators in the matrix model. We propose a general expression for the moments at any given order $n$. We also discuss orthogonal polynomials for spread as well as operator Krylov complexity. In the context of the Krylov operator growth we further compute the spectral function and the density of states. A careful analysis of the spectral function near the resonance reveals various IR divergences in addition to the physical collective modes. These collective modes appear to be exceptionally stable due to large mass gap. On the other hand, the UV of the theory appears to be extremely damped due to higher scattering rates. We carefully diagnose these IR divergences near resonance which reveals that these collective modes are in fact non-propagating and should be thought of as diffusion modes or relaxation modes with infinite relaxation time. We also calculate the Krylov variance and Krylov entropy for the matrix model and in particular at early time. The linear early time growth of the Krylov entropy confirms the onset of quantum chaos in the matrix model at large mass gap.}
\begin{document} 
\maketitle
\flushbottom
\section{Introduction and General Idea}
In recent years, the Krylov operator growth and the spread complexity of states \cite{Parker:2018yvk}-\cite{Caputa:2024vrn} have started to play a key role in understanding quantum information theory \cite{Nandy:2024evd}, and this has extended its domain in theoretical physics, especially in the context of chaos \cite{Hashimoto:2023swv}-\cite{Bhattacharjee:2024yxj} and holographic correspondence \cite{Caputa:2024sux}-\cite{Fatemiabhari:2025cyy}. See \cite{Baiguera:2025dkc}-\cite{Rabinovici:2025otw} for some nice comprehensible reviews.

The purpose of this paper is to extend and explore the notion of Krylov complexity in the context of BMN matrix models \cite{Berenstein:2002jq}-\cite{Huh:2024ytz} that are considered to be the massive deformation of the BFSS matrix model \cite{Banks:1996vh}. BMN (also known as the Plane Wave Matrix Model), as it stands today, offers an excellent testing laboratory for quantum gravity \cite{Dasgupta:2002hx}-\cite{Sugiyama:2002rs} that has been successfully lifted to the holographic framework \cite{Lin:2005nh}-\cite{Asano:2012zt}. 

The general idea is to start with a specific representation of the BMN model and reduce it to a simple Hamiltonian quantum mechanics following a suitable reduction ansatz \cite{Asano:2015eha}. In the present paper, we unify the $N=2$ and $N=4$ representations of the Fuzzy sphere vacuum within a single potential function \eqref{e2.6} that is characterized by parameters $k$, $l$ and $n$, which take different values for different Fuzzy sphere representations mentioned above.

The paper is divided mainly into two parts. In the first part, we discuss the moments and return amplitude in the context of spread complexity of states \cite{Balasubramanian:2022tpr}. The initial wavefunction (at $t=0$), with which we choose to begin with, is \emph{separable} as a product of two harmonic oscillator wave functions around their respective vacuum. In other words, the initial state represents two ``unentangled'' oscillators, in contrast to its thermo-field double counterpart \cite{Caputa:2025ozd}.  We compute first few moments ($\mathcal{M}_n$) in the matrix model, which has a generic characteristic at the large (mass) deformation $\mu \rightarrow \infty$ limit, which is $\mathcal{M}_n=\alpha_{(n)} \mu^n$, where the coefficients $\alpha_{(n)}$ are purely determined by the parameters of the matrix model\footnote{Contrary to the standard notation in the literature \cite{Balasubramanian:2022tpr}, here we use a slightly different notation, where we refer to $\mathcal{M}_n$ as the moment of order $n$ and the mass parameter as $\mu$.}. 

Next, we move on constructing the first few orthogonal polynomials \cite{Muck:2022xfc}-\cite{Muck:2024fpb} for the matrix model at strong coupling. We classify them for both $N=2$ and $N=4$ representations of the Fuzzy sphere vacuum. Looking at their structure for first few orders, we argue about their generic structure \eqref{e2.60} in the strong mass deformation limit.

In the second part of the paper, we discuss all the above issues in the context of the Krylov operator growth in the matrix model \cite{Roychowdhury:2026sgg}. For the operator growth we compute moments up to order four and find that only \emph{even} $n$ moments survive, revealing that one is looking at the time evolution of a Hermitian operator $\mathcal{O}(t)$. Our analysis reveals a generic structure for these moments similar to that in the example of the spread complexity. Next, we construct orthogonal polynomials pertinent to Krylov operator growth, which reveals an analogue feature as in the case of the spread complexity of states in the matrix model.  

Our analysis reveals that in the strong mass deformation, the wave function associated with the Krylov basis (both in the context of spread complexity of states and Krylov operator growth) can be schematically expressed as a polynomial of the form
\begin{align}
\label{ee1.1}
    \phi_{n}(t)|_{\mu \gg 1}=p_{n}^{(0)}+p_{n}^{(1)} \mu t+p_{n}^{(2)}\mu^2 t^2+p_{n}^{(3)} \mu^3 t^3+\mathcal{O}(\mu^4 t^4)
\end{align}
where $\mu t(<1)$ is kept fixed in the limit $\mu \rightarrow \infty$. In other words, the spread of the state or the operator is confined within the time domain $t<1$, depending on the coupling $\mu$. This would refer to a finite size of the Krylov subspace and the major contribution to the Krylov wavefunction would come from its early time evolution near $t \sim 0$. 

In the context of the Krylov operator growth analysis, we push this direction further by computing the spectral function and the resolvent \cite{Muck:2022xfc} for the matrix model at strong coupling $\mu$. As a first step, we calculate the Laplace transform of the wave function satisfying the Schrodinger equation. Our analysis leads to a natural generalization of these Laplace transform for arbitrary (even and odd) $n$. We identify the Laplace transform with the memory function \cite{Muck:2022xfc} which in turn is related to the \emph{monic} coefficients $\Delta_n$ of the matrix model in the large $\mu$ limit. Next, we compute the resolvent \cite{Muck:2022xfc}-\cite{Muck:2024fpb} using Laplace transform and categorize them for different representations of the Fuzzy sphere vacuum. 

The above building block, in particular the resolvent (which is identified as an operator-valued Green's function) serves as the key element to probe further into the many body dynamics of the matrix model at large mass gap. The imaginary component of the resolvent leads to the spectral function. The integral of the spectral function over the frequency domain leads to the density of states. Given the spectral function, we discuss various situations. As our analysis reveals, outside the resonance peak, the spectral function is well behaved and in particular vanishes at high frequencies. However, near IR, the spectral function exhibits divergences, and a special attention is needed. To understand these aspects more clearly, we investigate separately the behavior of the spectral function both at high and low frequency domain. As our analysis reveals, the vanishing of the spectral function in the UV corresponds to a high decay rate and a heavily damped configuration.

In deep IR, we classify different divergences appearing in the spectral function and isolate the physical contribution. The physical part of the spectral function corresponds to extremely stable collective modes with life time $\tau \propto 1/\omega^3$, which is different from the conventional Fermi liquid where $\tau \propto 1/\omega^2$. In other words, the low energy excitations in the matrix model are longer lived than conventional condensed matter systems. We identify this as an artifact of the large mass gap $\mu$ in the matrix model which prohibits scattering with high frequency modes. In other words, the large mass gap forbids the collectives (in the deep IR) to decay, making them exceptionally stable. 

We further present a proper diagnostic of the IR divergences appearing in the spectral function and propose possible methods to deal with these divergences while we treat the spectral function actually as a distribution under the integral over the full frequency domain. A closer analysis of the pole structure in the spectral function indicates that these collective modes are ``non-propagating'' and should be thought of as hydrodynamics, diffusive or highly collective non-Fermi liquid states. The non-Fermi liquid behavior is further confirmed from the power law behavior appearing in the pole of the spectral function at low frequencies.  We isolate this pole and propose a ``regularized'' spectral function which finally results in a finite density of states which does not depend on the representation of the Fuzzy sphere.

Finally, to complete and conclude our analysis, we calculate the Krylov variance \cite{Nandy:2024evd},\cite{Caputa:2021ori} and Krylov entropy \cite{Barbon:2019wsy}-\cite{Rabinovici:2020ryf} for the matrix model. As our analysis reveals, the Krylov variance scales quadratically with $t$ while the Krylov entropy scales linearly with $t$. The linear rate of growth of the Krylov entropy, which is controlled by the mass parameter $\mu$, further confirms the onset of quantum chaos \cite{Asano:2015eha} in BMN matrix model at large mass gap.  

Beyond the technical evaluation of Lanczos coefficients and orthogonal polynomials, our results hold direct significance for holography and quantum information dynamics in gauge theories. The Berenstein-Maldacena-Nastase (BMN) matrix model represents a deformation of the BFSS matrix model and serves as a tractable non-perturbative formulation of M-theory on a pp-wave background. Understanding how quantum information spreads, quantified via Krylov complexity, in such matrix quantum mechanics sheds light on the microscopic mechanisms of thermalization in holographic gravitational duals.

Consider the expansion \eqref{ee1.1} that has a natural consequence in the suppression of the Lanczos coefficients $b_n$ and hence the operator growth. The operator dynamics are effectively confined to early times $t < 1/\mu$. The large mass parameter $\mu$ traps the operator wavepacket near the origin of the Krylov chain, drastically slowing down information scrambling compared to the fast, linear growth of $b_n$ found in chaotic or conformal field theories. This has a direct holographic (or gravitational) counterpart as we explain below.

In holographic duals (such as $D0$-branes in M-theory on a pp-wave background), the mass parameter $\mu$ corresponds to the spacetime curvature/flux scale of the pp-wave. Large $\mu$ effectively turns off non-linear gravitational interactions and confines matrix excitations around fixed classical vacuum states (fuzzy spheres). Consequently, information stays localized in smaller subsectors of the operator space longer, preventing the exponential growth typical of fast scramblers (like CFTs or SYK models). Let us elaborate more on this.

In $11$-dimensional supergravity, the BMN plane-wave background is characterized by a constant 4-form flux $F_{+123} \propto \mu$. The mass parameter $\mu$ measures the spacetime curvature scale and the strength of the background magnetic flux. At large $\mu$, the background flux acts as a strong gravitational harmonic trap. In the bulk, supergravity excitations (gravitons and membranes) experience an effective potential proportional to $\mu^2 r^2$, preventing them from freely propagating to large radial distances or dispersing across the string geometry.

In matrix quantum mechanics, the classical vacua of the BMN model correspond to non-commutative fuzzy spheres $\sum_{i=1}^3 [X_i, X_j] = i \mu \epsilon_{ijk} X_k$. At $\mu \to \infty$, the large mass gap forces the quantum fluctuations to remain tightly bound around these classical fuzzy sphere configurations. In CFTs or SYK-like models (unconstrained fast scramblers), operator excitations mix rapidly across the entire Hilbert space via unsuppressed non-linear terms, yielding $b_n \sim \alpha n$ and exponential operator growth. In BMN at large $\mu$, the off-diagonal matrix couplings that connect different fuzzy sphere topological flux sectors are strongly suppressed by $1/\mu$. This confines operator dynamics to a small, localized subsector of the Krylov chain for extended time intervals ($t < 1/\mu$).

Furthermore, our extraction of the spectral function and density of states reveals a critical physical phenomenon: the appearance of IR divergences near resonance corresponding to non-propagating collective modes. At large mass gap $\mu \gg 1$, these modes represent long-lived relaxation or diffusion modes rather than propagating excitations, signaling a distinct phase in the operator dynamics of mass-deformed gauge theories. Because these collective modes have non-zero weight but infinite relaxation times in the strict large-$\mu$ limit, they act as long-lived memory channels in the operator space. This demonstrates why mass-deformed matrix models exhibit slow thermalization and delayed information scrambling. This phenomenon is analogous to condensed matter systems with large energy gaps (e.g., topological insulators or gapped superconductors), where the bulk exhibits a large mass gap while gapless, highly stable non-decaying collective modes persist at zero frequency. Finally, the early-time linear growth of Krylov entropy rigorously establishes the onset of scrambling in a system with discrete flux sectors (fuzzy sphere vacua). This positions Krylov complexity as a robust, scheme-independent probe of quantum chaos in gauge theories with non-trivial vacuum structure, bridging quantum information theory with string and/or M-theory dynamics.

The organization of the rest of the paper is as follows. We discuss moments and return amplitude for spread complexity of states in the matrix model in Section \ref{sec2}. We also discuss orthogonal polynomials and their general structure. In Section \ref{sec3}, we discuss moments and the basic structure of orthogonal polynomials for the Krylov operator growth. In Section \ref{sec4}, we use the notion of Krylov operator growth to compute the spectral density function for the BMN matrix model and perform a close analysis of the IR divergences and discuss its physical consequences on the low energy spectrum of the theory. In Section \ref{sec5}, we perform a detailed calculation on the Krylov variance and the Krylov entropy of the matrix model at large mass gap. Finally, we conclude in Section \ref{sec6} and list a set of interesting problems. 

\section{Moments and return amplitude for spread complexity of states}
\label{sec2}
The organization of this Section is the following. We begin by reviewing the basics of the Fuzzy sphere model \cite{Asano:2015eha} and the derivation of the first few Lanczos coefficients $a_n$ and $b_n$ in the limit of large mass deformation $\mu \rightarrow \infty$. Next, we construct the return amplitude and the moments \cite{Balasubramanian:2022tpr} at strong coupling and obtain their most generic expressions.
\subsection{Preliminaries}
The BMN matrix model \cite{Berenstein:2002jq} is defined by the following action\footnote{Throughout this paper, we choose to work with natural units where we set $\hbar =c=l=1$.} \cite{Asano:2015eha}
\begin{align}
\label{e1.1}
    S_{PWMM}&=\frac{1}{g^2}\int dt \text{Tr}\Big[ \frac{1}{2}(D_t X^r)^2+\frac{1}{4}[X^r,X^s]^2 -\frac{1}{2}\Big(\frac{\mu}{3}\Big)^2X^2_i \nonumber\\
    &-\frac{1}{2}\Big(\frac{\mu}{6}\Big)^2X^2_a -\frac{i\mu}{3}\epsilon_{ijk}X^i X^j X^k\Big]+\text{Fermions}.
\end{align}

The above action \eqref{e1.1} is obtained starting with $\mathcal{N}=4$ in $R\times S^3$ and thus following a reduction along the three sphere. Here $\mu$ is the mass deformation parameter, which is considered to be large enough in our computation. Here, $X^r (r=1,\cdots , 9)$ are the $U(N)$ gauge invariant matrices that satisfy the following set of equations \cite{Asano:2015eha}
\begin{align}
\label{e2.2}
    &\ddot{X}^i+[X^r,[X^r,X^i]]+\Big(\frac{\mu}{3}\Big)^2 X^i+i \mu \epsilon^{ijk}X_j X_k =0\\
    \label{e2.3}
    &\ddot{X}^a+[X^r,[X^r,X^a]]+\Big(\frac{\mu}{6}\Big)^2 X^a =0
\end{align}
along with the Gauss law constraint
\begin{align}
\label{e2.4}
    [X^r,\dot{X^r}]=0.
\end{align}

Here, $i,j=1,2,3$ are the $SU(2)$ indices and $a,b=4,\cdots ,9$ are the $SO(6)$ indices. In what follows, we consider several consistent reductions of the matrix model that are compatible with \eqref{e2.2}-\eqref{e2.4}. The first example we consider goes under the name of the Fuzzy sphere model \cite{Asano:2015eha}, characterized by the following Hamiltonian density
\begin{align}
\label{e2.5}
    \mathcal{H}=\frac{1}{2}(p_x^2+p_y^2)+V(x,y).
\end{align}

The potential function can be schematically expressed as \cite{Roychowdhury:2026igc}
\begin{align}
\label{e2.6}
    V(x,y)=\frac{\mu^2}{2}(k x^2+y^2)+\frac{1}{2}(x^4+y^4)+l x^2 y^2-\mu y^3-n \mu x^2 y.
\end{align}

\paragraph{Important note.} Here, we introduce certain set of levels (or parameters) to classify different representations of the matrix model. For example, setting $k=\frac{1}{4}$, $l=1$ and $n=0$ we have a matrix model configuration known as the pulsating fuzzy sphere (PFS) \cite{Asano:2015eha}, which corresponds to the $N=2$ representation of the $SU(2)$ algebra. On the other hand, setting the parameters as $k=1$, $l=3$, $n=3$ correspond to the integrable fuzzy sphere model (IFS), which is a $N=4$ representation of the $SU(2)$ algebra \cite{Asano:2015eha}. \\\\
\uline{\textbf{Constructing the Krylov basis:}}\\\\
Our next task is to construct the Krylov basis elements $\{\ket{K_n}\}$ that are orthonormal $\braket{K_n|K_m}=\delta_{nm}$ and satisfy the Krylov chain criterion \cite{Balasubramanian:2022tpr}
\begin{align}
    \mathcal{H} \ket{K_n}=a_n \ket{K_n}+b_n \ket{K_{n-1}}+b_{n+1}\ket{K_{n+1}}.
\end{align}

Here, $a_n$ and $b_n$ are the Lanczos coefficients, which can be expressed as
\begin{align}
    a_n=\braket{K_n|\mathcal{H}|K_n}~;~b_n = \braket{K_{n-1}|\mathcal{H}|K_n}.
\end{align}

In quantum mechanics, a general state $\ket{\Psi(t)}$ is expressed by unitary time evolution
\begin{align}
\label{e2.9}
     \ket{\Psi(t)}=\sum_{n=0}^{\infty}\frac{(-it)^n}{n!}\mathcal{H}^n \ket{\Psi_0}=\sum_{n=0}^{\infty}\frac{(-it)^n}{n!}\ket{\Psi_n}
\end{align}
which could be thought of as an expansion in a basis $\{\ket{\Psi_n}\}=\{\mathcal{H}^n\ket{\Psi_0}\}$.

The state $\ket{\Psi(t)}$ can be expanded in the Krylov basis as
\begin{align}
\label{e2.10}
 \ket{\Psi(t)}=\sum_{n=0}^{\infty}\psi_n(t)\ket{K_n}   
\end{align}
where the coefficients $\psi_n(t)$ are fixed by the Krylov chain condition and satisfy an equation
\begin{align}
\label{e2.11}
     i \partial_t \psi_n (t)=a_n \psi_n (t)+b_{n+1}\psi_{n+1}(t)+b_n \psi_{n-1}(t).
\end{align}

Spread complexity is defined as the average location on the Krylov chain \cite{Balasubramanian:2022tpr}
\begin{align}
    C(t)=\sum_{n=0}^\infty n | \psi_n(t)|^2.
\end{align}

After analytically solving the first few orthonormal wave functions $\psi_n(t)(n=0,1,2)$, the LO contribution to the spread complexity can be expressed as
\begin{align}
    C(t)|_{t\sim 0}=\zeta \mu^2 t^2+\mathcal{O}(\mu^4 t^4)
\end{align}
where the entity $\zeta$ is purely fixed by the Lanczos coefficients\footnote{See Section \ref{sec2.3} for a detailed derivation.} 
\begin{align}
\label{ee2.14}
    \zeta=\frac{b^2_1}{\mu^2}+\text{sub-leading corrections}
\end{align}
where sub-leading corrections are found to be proportional to $b_2^2 / (a_0 + a_1)^2$.

Our first step would be to construct the Krylov basis elements using the Gram-Schmidt procedure. We propose an expansion of the following form \cite{Roychowdhury:2026vzq}
\begin{align}
\label{e2.15}
    \ket{K_{n+1}}=\ket{\Psi_{n+1}}-c_n \ket{K_n}-d_n \ket{K_{n-1}}~;~n=0,1,2,\cdots
\end{align}
where the coefficients $c_n$ and $d_n$ are obtained from the orthonormality criteria.

Setting $n=0$, we have the first Krylov chain condition (with $\ket{K_{-1}}=0$)
\begin{align}
\label{e2.16}
 \ket{K_{1}}=\ket{\Psi_{1}}-c_0 \ket{K_0}. 
\end{align}

We choose $\ket{K_0}=\ket{\Psi_0}$ as the initial state such that $\braket{K_0|K_0}=1$. Following the arguments of \cite{Amore:2024ihm}-\cite{Huh:2024ytz}, we choose the initial configuration as the \emph{localized} harmonic oscillators around their respective minima $x \sim 0$ and $y \sim 0$ of the potential. This can be argued following a scaling such that $x \rightarrow 0$ and $y \rightarrow 0$ while keeping $\mu x$ and $\mu y$ as finite. This would result in the lowest energy configuration of the system in the presence of an effective low energy potential which appears to be a combination of two harmonic oscillator potentials\footnote{Notice that in the expansion \eqref{e2.17} we have considered $\mu x$ and $\mu y$ to be \emph{fixed} $\sim\mathcal{O}(1)$ in the limit $x,y \rightarrow 0$ while setting $\mu$ to be large enough. As a result, the first sub-leading correction in \eqref{e2.6} appears to be $\mu y^3 \sim \mathcal{O}(y^2)$ in the limit $y \sim 0$ and is therefore smaller as compared to the harmonic potential.}
\begin{align}
\label{e2.17}
    V(x,y)=\frac{\mu^2}{2}(k x^2+y^2)+\mathcal{O}(y^2).
\end{align}

Following the above line of arguments, we choose the ground state wavefunction, following a suitable rescaling of the coordinates as \cite{Roychowdhury:2026sgg}
\begin{align}
\label{e2.18}
    &\ket{K_0}=\ket{\Psi_0}=\int dx dy \Psi_0(x,y)\ket{x,y}=\ket{\Psi_0(x)}\otimes \ket{\Psi_0(y)}\\
    & \Psi_0(x,y)=\sqrt{\frac{2 \mu}{\pi}}e^{-\mu(x^2+y^2)}=\Psi_0(x)\Psi_0(y)~;~\ket{x,y}=\ket{x}\otimes \ket{y}.
    \label{e2.19}
\end{align}

Here, $\ket{x,y}$ is the eigen basis in the position representation which satisfies the orthonormal criteria, that is, $\braket{x,y|x',y'}=\delta(x-x')\delta (y-y')$. At this stage, it is worth emphasizing that the initial state \eqref{e2.19} corresponds to a direct product of two harmonic oscillator states localized around their respective vacuum. In other words, our initial choice corresponds to two \emph{unentangled} oscillators in their respective ground states. 

The state $\ket{\Psi_1}$ can be obtained using \eqref{e2.5} and \eqref{e2.18}
\begin{align}
    &\ket{\Psi_1}=\mathcal{H}\ket{\Psi_0}=\int dx dy\Phi_1(x,y)\Psi_0(x,y)\ket{x,y}\nonumber\\
   & \Phi_1(x,y)=2 \mu (1-\mu(x^2+y^2))+V(x,y).
\end{align}

Taking an inner product with $\ket{K_0}$ on both sides of \eqref{e2.16} and setting $\braket{K_0|K_1}=0$, we find the coefficient $c_0$
\begin{align}
\label{e2.21}
    c_0=\braket{K_0|\Psi_1}=\frac{1}{16 \mu ^2}\Big[2 (k+9) \mu ^3+l+3 \Big]=\frac{\mu}{8} (k+9) +\mathcal{O}(\mu^{-2}).
\end{align}

The state $\ket{K_1}$ should be normalized, that is, $\ket{K_1}\rightarrow \frac{1}{\sqrt{N_1}}\ket{K_1}$, where the normalization constant is given by
\begin{align}
\label{e2.22}
    N_1&=\braket{K_1|K_1}=\frac{1}{64 \mu ^4}\Bigg[2 ((k-8) k+25) \mu ^6\nonumber\\
    &+\mu ^3 (2 k (l+3)-14 l+3 n (n+2)-27)+2 (l (l+3)+6) \Bigg].
\end{align}

Similarly for $n=1$, we have the following Krylov state
\begin{align}
    \ket{K_2}=\ket{\Psi_2}-c_1\ket{K_1}-d_1 \ket{K_0}.
\end{align}

The state $\ket{\Psi_2}$ is given by
\begin{align}
   &\ket{\Psi_2} =\mathcal{H}^2\ket{\Psi_0}=\int dx dy \Phi_2(x,y)\Psi_0(x,y)\ket{x,y}\nonumber\\
   & \Phi_2(x,y)=\frac{1}{2}\mathcal{F}(x,y)+V(x,y)\Phi_1(x,y).
   \label{e2.24}
\end{align}

The function above in \eqref{e2.24} can be expressed as
\begin{align}
  &\mathcal{F}(x,y)= -2 (l+3) \left(x^2+y^2\right)+ 2 \mu  \left(5 \left(2 l x^2 y^2+x^4+y^4\right)+(n+3) y\right)\nonumber\\
  &-\mu ^2 \left(k+2 \left(x^2 \left(2 l y^4+8 n y+y^4\right)+(2 l+1) x^4 y^2+x^6+y^6+8 y^3\right)-15\right)\nonumber\\
  &+2 \mu ^3 \left(x^2 \left(3 k+2 (n+1) y^3-16\right)+2 n x^4 y+y^2 \left(2 y^3-13\right)\right)\nonumber\\
  &-\left.2 \mu ^4 \left(x^2+y^2\right) \left((k-4) x^2-3 y^2\right)\right).
\end{align}

Taking the inner product with $\ket{K_0}$ and setting $\braket{K_o|K_2}=0$, we find
\begin{align}
\label{e2.26}
    d_1&=\braket{K_0|\Psi_2}=\frac{1}{256 \mu ^4}\Bigg[4 (k (3 k+2)+131) \mu ^6\nonumber\\
    &+4 \mu ^3 (3 k (l+3)-5 l+3 n (n+2))+9 l^2+30 l+57 \Bigg].
\end{align}

On a similar note, taking inner product with $\ket{K_1}$ and setting $\braket{K_1|K_2}=0$, we find
\begin{align}
\label{e2.27}
    c_1 N_1 &= \braket{K_1|\Psi_2}\nonumber\\
    &=\frac{1}{512 \mu ^6}\Bigg[3 (l+3) (l (9 l+10)+37)+2\mu^3\Big(3 k (l (5 l+16)+31)-25 l^2\nonumber\\
    &+141+l (33 n (n+2)-69)+54 n (n+2)\Big)+2\mu^6 \Big(9 k^2 (l+4)-38 k l-103 l\nonumber\\
    &-354+3 k (n+3) (7 n-13)-2 n (8 n+39)\Big) +4\mu ^9 (k (k (3 k-7)-79)+443) \Bigg].
\end{align}

Finally, we compute the normalization constant for the state $\ket{K_2}$, which yields
\begin{align}
\label{e2.28}
    N_2=\braket{K_2|K_2}=\int dx dy \Psi_0^2 (x,y)\Phi^2_2(x,y)-N_1 c_1^2 -d_1^2.
\end{align}

The above integral \eqref{e2.28} can be performed, and it produces a large output which we prefer not to quote here. However, in the limit $\mu \rightarrow \infty$ it simplifies a lot to yield 
\begin{align}
    N_2&= \frac{\mu^4}{512 ((k-8) k+25)}\Big[3 k \Big(k \Big(k (k ((k-24) k+267)-1640)\nonumber\\
    &+5811\Big)-11544\Big)+33051 \Big]+\mathcal{O}(\mu).
\end{align}

In summary, we have the following normalized Krylov states
\begin{align}
\label{e2.30}
  &\ket{K_0}=\ket{\Psi_0}\\
  \label{e2.31}
  &\ket{K_1}=\frac{1}{\sqrt{N_1}}\Big[ \ket{\Psi_1}-c_0\ket{K_0}\Big]\\
  &\ket{K_2}=\frac{1}{\sqrt{N_2}}\Big[ \ket{\Psi_2}-c_1\ket{K_1}-d_1 \ket{K_0}\Big].
  \label{e2.32}
\end{align}
which will be used in the next section to compute the return amplitude.
\subsection{Return amplitude and Lanczos coefficients}
\label{sec2.2}
To compute the return amplitude $\mathcal{R}(t)$, we first notice the following expansions that readily follow from \eqref{e2.9} and \eqref{e2.10} respectively
\begin{align}
\label{e2.33}
     & \ket{\Psi(t)}=\ket{\Psi_0}-it \ket{\Psi_1}+\cdots\\
    & \ket{\Psi(t)}=\psi_0(t)\ket{K_0}+\psi_1(t)\ket{K_1}+\cdots. 
    \label{e2.34}
\end{align}

It is instructive to highlight the dimensionless combination that governs our perturbative regime. 
By scaling the coordinates as $\mu x, \mu y \sim \mathcal{O}(1)$ near the effective minimum, the leading quadratic potential term scales as $V \sim \mu^2$, which fixes the relevant energy scale of the localized initial ground state $\ket{\Psi_0}$ to be proportional to $\mu$. 
Consequently, the natural dimensionless parameter for time evolution in this setup is the product $\mu t$. 
In the large-mass regime $\mu \gg 1$, the Lanczos coefficients scale linearly as $b_1(\mu) \propto \mu$, meaning that the short-time expansion $t \to 0$ is systematically controlled by expanding in powers of $\mu t \ll 1$. 
This dimensionless organizing parameter ensures the consistency of our early-time approximation alongside the truncation of the potential at $\mathcal{O}(\mu^2)$.

Taking the inner product with $\ket{K_1}$, we find
\begin{align}
\label{ee2.35}
    \psi_1(t)=-\frac{it}{\sqrt{N_1}}\Big[ \braket{\Psi_1|\Psi_1}-c^2_0\Big]=-it b_1(\mu)+\mathcal{O}(\mu^2t^2).
\end{align}

A straightforward computation reveals the above entity as
\begin{align}
    &b_1 (\mu)= \frac{1}{8 \mu^2}\Big[ 2 ((k-8) k+25) \mu ^6+\mu ^3 (2 k (l+3)-14 l+3 n (n+2)-27)\nonumber\\
    &+2 (l (l+3)+6)\Big]^{1/2}=\frac{ \mu }{4 \sqrt{2} } \sqrt{\left(k^2-8 k+25\right) }+\mathcal{O}(\mu^{-2}).
\end{align}

Next, we compute the wavefunction $\psi_0(t)$, which satisfies an equation 
\begin{align}
    i \partial_t \psi_0 (t) = a_0 \psi_0(t)+b_1 \psi_1 (t)
\end{align}
which has a solution of the form
\begin{align}
\label{e2.38}
 \psi_0(t)= \frac{b_1^2 }{a^2_0}(-1+i a_0 t)+\Big(1+\frac{b^2_1}{a^2_0} \Big) e^{-i a_0 t}. 
\end{align}

Notice that considering the fact that $a_0,b_1 \sim \mu$ at strong mass deformation, one can schematically express the wave functions corresponding to $n-0,1$ as
\begin{align}
    \psi_0(t)=p_0^{(0)}+p_0^{(1)} \mu t+p_0^{(2)}e^{-i\mu t}~,~\psi_1(t)=p_1^{(1)}\mu t +\mathcal{O}(\mu^2 t^2)
\end{align}
which is typically of the polynomial form \eqref{ee1.1}, which is subject to the fact $\mu t <1$ in the limit of the large mass gap $\mu \rightarrow \infty$. In other words, the spread of the wavefunction \eqref{e2.10} is mostly localized in a small time domain, indicating a finite size of the Krylov subspace.

Notice that $\psi_0(t=0)=1$, which ensures that $\braket{\Psi(t)|\Psi(t)}|_{t=0}=1$. The return (or survival) amplitude \cite{Balasubramanian:2022tpr} is defined as the complex conjugate of \eqref{e2.38}, that is,
\begin{align}
\label{e2.39}
\mathcal{R}(t)=\braket{\Psi(t)|\Psi(0)}=\sum_{n=0}^{\infty}\psi^\ast_n(t)\braket{K_n|K_0}=\psi_0^\ast (t).
\end{align}

Notice that the return amplitude \eqref{e2.39} (and hence the moments computed below) are valid only up to certain order in time $t$ as an expansion over the Krylov lattice. This is evident from the truncation in the series \eqref{e2.33}-\eqref{e2.34}. In other word, the wave functions $\psi_n(t)$ would be corrected at higher order in time $t$. However, the nice thing about these constructions is that the generic structure for the first few terms in the expression for the $n$-th moment could be read out by assuming the fact that the return amplitude \eqref{e2.39} is perturbatively corrected at higher order in the expansion in time $t$. Similar remarks also hold for the Krylov operator growth which we discuss the next section.

The $n$-th moment is defined as the time derivative of \eqref{e2.39}
\begin{align}
    \mathcal{M}_n=\frac{d^n}{dt^n}\mathcal{R}(t)\Big|_{t=0}=\braket{K_0|(i\mathcal{H})^n|K_0}.
\end{align}

For $n=1$, we find the first moment
\begin{align}
    \mathcal{M}_1(\mu)=\frac{d \psi^\ast_0(t)}{dt}\Big|_{t=0}=i a_0(\mu)=\frac{i\mu}{8} (k+9) +\mathcal{O}(\mu^{-2})=\alpha_{(1)}\mu+\mathcal{O}(\mu^{-2}).
\end{align}

Notice that the leading coefficient $\alpha_{(1)}$ is characterized by the parameter $k$, which depends of the Fuzzy sphere model that one chooses to work with
\begin{equation}
\label{e2.42}
 \mathcal{M}_1(\mu)=
    \begin{cases}
      \frac{37i}{32}\mu & \text{PFS}~ (k=1/4)\\
      \frac{5i}{4}\mu & \text{IFS}~(k=1).
    \end{cases}       
\end{equation}

On a similar note, we find for $n=2$, we find the second moment
\begin{align}
    &\mathcal{M}_2(\mu)=\frac{d^2 \psi^\ast_0(t)}{dt^2}\Big|_{t=0}=-a^2_0-b^2_1=
    \alpha_{(2)}\mu^2\nonumber\\
    &\alpha_{(2)}=-\frac{1}{64} \Big[ k (3 k+2)+131\Big].
\end{align}
   
Clearly, like the first moment, the second moment also depends on the choice of the parameter $k$ of the matrix model, that is,
\begin{equation}
\label{e2.44}
 \mathcal{M}_2(\mu)=
    \begin{cases}
      2.05762\mu^2 & \text{PFS}~ (k=1/4)\\
      2.125\mu^2 & \text{IFS}~(k=1).
    \end{cases}       
\end{equation}

The above calculation can be further extended for higher order Krylov lattice points which would produce moments at higher order that are generated at a later time evolution. Looking at these expressions, we propose a generic expression for the $n$-th moment
\begin{align}
\label{e2.45}
    \mathcal{M}_n(\mu)=\alpha_{(n)}\mu^n
\end{align}
where the coefficient $\alpha_{(n)}$ depends on the representation of the Fuzzy sphere vacuum.

Considering the fact that \eqref{e2.38} would be perturbatively corrected at higher order in the time evolution, we propose the generic structure for the moments (for $n>1$)
\begin{align}
\label{e2.46}
    \mathcal{M}_{n}(\mu)=\frac{d^n}{dt^n}\Big(1+\frac{b_1^2}{a_0^2} \Big)e^{ia_0 t}\Big|_{t=0}+f(a_n,b_n)=i^na_0^n\Big(1+\frac{b_1^2}{a_0^2} \Big)+f(a_n,b_n)
\end{align}
where the remaining contribution $f(a_n,b_n)$ has to be determined by knowing higher order terms in the time expansions \eqref{e2.33} and \eqref{e2.34}.

Using the expressions for $a_0(\mu)$ and $b_1(\mu)$, one can deduce the generic structure for the first term in \eqref{e2.46} and read out the coefficient $\alpha_{(n)}$ for the $n$-th moment (for $n> 1$)
\begin{align}
    \alpha_{(n)}=\frac{i^n}{2^{3n}}(k+9)^n \Big[ 1+\frac{2(k^2-8k+25)}{(k+9)^2}\Big]+f^{(n)}(a_n,b_n)
\end{align}
which clearly reveals the dependence of the moments $\mathcal{M}_n$ on the choice ($k$) of the Fuzzy sphere vacuum, as is already evident from \eqref{e2.42} and \eqref{e2.44}.
\subsection{Early-time spread complexity and universal $t^2$ growth}
\label{sec2.3}
To clarify the early-time behavior of the spread complexity $C(t)$, we analyze the Taylor expansion of the wavefunctions $\{\psi_n(t)\}$ near $t \to 0$. From the Schrödinger-like equation governing the Krylov chain condition \eqref{e2.11}
\begin{equation}
i \partial_t \psi_n(t) = a_n \psi_n(t) + b_{n+1} \psi_{n+1}(t) + b_n \psi_{n-1}(t) \quad \text{with} \quad \psi_n(0) = \delta_{n0},
\label{eq:krylov_schrodinger}
\end{equation}
the exact short-time behavior of lowest Krylov modes follow directly from \eqref{ee2.35} and \eqref{e2.38}
\begin{align}
\psi_0(t) &= \frac{b_1^2}{a_0^2}\left(-1 + i a_0 t\right) + \left(1 + \frac{b_1^2}{a_0^2}\right) e^{-i a_0 t} = 1 - i a_0 t - \frac{1}{2}\left(a_0^2 + b_1^2\right) t^2 + \mathcal{O}(t^3), \label{eq:psi0_expansion} \\
\psi_1(t) &= -i b_1 t + \mathcal{O}(t^2), \label{eq:psi1_expansion} \\
\psi_{n}(t) &= \mathcal{O}(t^n) \quad \text{for } n \ge 2.
\end{align}

Evaluating the probabilities $p_n(t) = |\psi_n(t)|^2$ up to second order in time yields
\begin{align}
p_0(t) &= |\psi_0(t)|^2 = 1 - b_1^2 t^2 + \mathcal{O}(t^4), \\
p_1(t) &= |\psi_1(t)|^2 = b_1^2 t^2 + \mathcal{O}(t^4), \\
p_n(t) &= \mathcal{O}(t^4) \quad \text{for } n \ge 2.
\end{align}

By definition, the spread complexity is the expectation value of the position operator along the Krylov chain
\begin{equation}
C(t) = \sum_{n=0}^{\infty} n |\psi_n(t)|^2 = 1 \cdot p_1(t) + 2 \cdot p_2(t) + \dots = b_1^2 t^2 + \mathcal{O}(t^4).
\label{eq:spread_complexity_exact}
\end{equation}

Equation \eqref{eq:spread_complexity_exact} demonstrates that the early-time quadratic growth of spread complexity is universally governed by $b_1^2 t^2$, independent of higher Lanczos coefficients $(a_1, b_2, \dots)$.

In our model, under the large mass deformation limit $\mu \gg 1$, $b_1(\mu)$ scales as $b_1 \sim \mu$. Consequently, the early-time growth rate equation can be written as
\begin{equation}
C(t)\Big|_{t \to 0} = \zeta \mu^2 t^2 + \mathcal{O}(\mu^4 t^4), \quad \text{where} \quad \zeta \equiv \frac{b_1^2}{\mu^2}.
\label{eq:corrected_complexity}
\end{equation}

The sub-leading terms proportional to $b_2^2 / (a_0 + a_1)^2$ in \eqref{ee2.14} correspond to higher-order perturbative corrections in the parameter $(\mu t)^4$ or truncation effects from intermediate coupling scales. Restoring the exact Krylov basis dynamics ensures that Eq.~\eqref{eq:corrected_complexity} strictly obeys the universal early-time expansion property of state spread complexity.
\subsection{Remarks on orthogonal polynomials}
Next, we discuss the notion of orthonormal polynomials \cite{Muck:2022xfc}-\cite{Muck:2024fpb} for BMN matrix models at strong mass deformation. The $n$-th Krylov basis element can be expressed as
\begin{align}
\label{e2.47}
    \ket{K_n}=\mathcal{P}_n(\mathcal{H})\ket{K_0}
\end{align}
which is equivalent to the recursion relation of the form \cite{Caputa:2025ozd}
\begin{align}
\mathcal{H}\mathcal{P}_n(\mathcal{H})=a_n\mathcal{P}_n(\mathcal{H})+b_n\mathcal{P}_{n-1}(\mathcal{H})+b_{n+1}\mathcal{P}_{n+1}(\mathcal{H}).
\end{align}

Setting $n=0$, we first notice that 
\begin{align}
    \ket{K_0}=\mathcal{P}_0(\mathcal{H})\ket{K_0}~;~\Rightarrow \mathcal{P}_0(\mathcal{H})=1.
\end{align}

Next, we set $n=1$, which yields 
\begin{align}
    \ket{K_1}=\mathcal{P}^{(k)}_1(\mathcal{H})\ket{K_0}
\end{align}
where the level $k$ refers to different representations \cite{Asano:2015eha} of the matrix model.

Comparing with \eqref{e2.31}, we find the following
\begin{align}
\label{e2.51}
    \mathcal{P}^{(k)}_1(\mathcal{H})=\frac{1}{\sqrt{N_1}}(\mathcal{H}^{(k)}-a_0)
\end{align}
where we identify $a_0 = c_0=\braket{K_0|\mathcal{H}|K_0}$, see Eq. \eqref{e2.21}.

Finally, taking $\mu$ to be large enough, one finds
\begin{align}
     \mathcal{P}^{(k)}_1(\mathcal{H})=\frac{4 \sqrt{2} }{\mu  \sqrt{k^2-8 k+25 }}\Big(\mathcal{H}^{(k)}-\frac{\mu}{8} (k+9) \Big).
\end{align}

Choosing different representations of the Fuzzy sphere vacuum, this simply yields
\begin{equation}
  \mathcal{P}_1(\mathcal{H})=
    \begin{cases}
      \frac{16 \sqrt{\frac{2}{41}}}{3 \mu }(\mathcal{H}_{PFS}-\frac{37}{32}\mu) & \text{PFS}~ (k=1/4)\\
     \frac{4}{3 \mu }(\mathcal{H}_{IFS}-\frac{5}{4}\mu) & \text{IFS}~(k=1).
    \end{cases}       
\end{equation}

Let us now construct the polynomial at higher order in the Hamiltonian $\mathcal{H}$. In order to find the polynomial for $n=2$, we propose the following
\begin{align}
\label{e2.54}
    \mathcal{P}^{(k)}_2(\mathcal{H})=\frac{1}{\sqrt{N_2}}\Big[(\mathcal{H}^{(k)} -\gamma_1) (\mathcal{H}^{(k)}-\gamma_2)+\gamma^2_2\Big].
\end{align}

Using \eqref{e2.54} and considering the definition \eqref{e2.47}, we find the following Krylov state
\begin{align}
    \ket{K_2}=\frac{1}{\sqrt{N_2}}\Bigg[\ket{\Psi_2}-(\gamma_1+\gamma_2)\ket{\Psi_1}+\gamma_2 (\gamma_1+\gamma_2)\ket{K_0} \Bigg].
\end{align}

Comparing with \eqref{e2.32}, we notice the following
\begin{align}
\label{e2.57}
    \gamma_1 = c_1-c_0+\frac{d_1}{c_1}~;~\gamma_2=c_0-\frac{d_1}{c_1}.
\end{align}

Clearly, in the strict large $\mu$ limit one has a simplification of the form
\begin{align}
   &\mathcal{P}^{(k)}_2(\mathcal{H})=  \frac{1}{\sqrt{N_2}}\Big[ \mathcal{H}^{2(k)}+\alpha_1(k)\mu\mathcal{H}^{(k)}+\alpha_2(k)\mu^2\Big]\\
   &\alpha_1(k)=-\frac{1}{4} \left(17+3k-\frac{18 (k-1)}{(k-8) k+25}\right)\\
   &\alpha_2(k)=\frac{(k (k (k (3 k+62)-474)+462)+4699) }{64 ((k-8) k+25)}.
\end{align}

Like before, the orthogonal polynomial at second order depends on the choice of the Fuzzy sphere vacuum. In particular, one finds the following expressions for the polynomial
\begin{equation}
\label{e2.61}
  \mathcal{P}_2(\mathcal{H})=
    \begin{cases}
     \frac{0.68}{ \mu ^2}\Bigg[\mathcal{H}^2_{PFS}-4.58 \mu\mathcal{H}_{PFS}+3.24 \mu^2\Bigg] & \text{PFS}~ (k=1/4)\\
    \frac{0.88}{ \mu ^2}\Bigg[\mathcal{H}^2_{IFS}-5 \mu \mathcal{H}_{IFS}+4.12 \mu^2\Bigg] & \text{IFS}~(k=1).
    \end{cases}       
\end{equation}

Here, the subscripts correspond to different Hamiltonian operators $\mathcal{H}$ that correspond to different representation of the Fuzzy sphere vacuum. Also it is interesting to notice that different $k$ values correspond to different numerical pre-factor which distinguishes these polynomials \eqref{e2.61} from one representation of the Fuzzy sphere to the other.

Notice that the above numbers clearly characterize the choice of the Fuzzy sphere vacuum, that is, $N=2$ or $N=4$ representation of the $SU(2)$ algebra. We propose a general expression of these polynomials characterizing the matrix model at large $\mu$
\begin{align}
\label{e2.60}
    \mathcal{P}^{(k)}_n(\mathcal{H})\Big|_{\mu \gg 1}=\frac{c^{(k)}}{\mu^n}\Big[\mathcal{H}^{n(k)} +\xi^{(k)}_{n-1}\mu\mathcal{H}^{n-1(k)}+\xi^{(k)}_{n-2}\mu^2 \mathcal{H}^{n-2(k)} +\cdots +\xi^{(k)}_{0} \mu^{n}\Big]
\end{align}
where $n\geq 1$ and $k$ label the choice of vacuum of the Fuzzy sphere. All remaining coefficients (for a given order $n$ of the polynomial) $c^{(k)}$, $\xi_n^{(k)}$ are fixed by choosing this vacuum. As a trivial illustration, one should notice that for $n=1$ we have $\xi_0^{(k)}=-a_0$.\\\\
\uline{\textbf{Orthogonality of the polynomials:}}\\\\  
The next thing we would like to ensure is the orthogonality of the above polynomials \eqref{e2.51} and \eqref{e2.54} by taking their explicit expressions. As a trivial check, one should notice
\begin{align}
    \braket{K_0|\mathcal{P}^2_0(\mathcal{H})|K_0}=1.
\end{align}

Next, we compute the following entity
\begin{align}
    \braket{K_0|\mathcal{P}_0(\mathcal{H})\mathcal{P}_1(\mathcal{H})|K_0} =\frac{1}{\sqrt{N_1}}\Big[ \braket{K_0|\mathcal{H}|K_0}-a_0\Big]=0.
\end{align}

Similarly, one can show
\begin{align}
  \braket{K_0|\mathcal{P}_0(\mathcal{H})\mathcal{P}_2(\mathcal{H})|K_0}&= \frac{1}{\sqrt{N_2}}\Big[\braket{K_0|\Psi_2}-(\gamma_1+\gamma_2)\braket{K_0|\Psi_1} +\gamma_2(\gamma_1+\gamma_2)\Big]\nonumber\\
  &=\frac{1}{\sqrt{N_2}}\Big[ \braket{K_0|\Psi_2}-c_1\braket{K_0|\Psi_1}+c_0 c_1-d_1\Big].
\end{align}

Using the fact that $\braket{K_0|\Psi_2}=d_1$ (see Eq. \eqref{e2.26}) and $\braket{K_0|\Psi_1}=c_0$ (see Eq. \eqref{e2.21}), it is straightforward to show
\begin{align}
  \braket{K_0|\mathcal{P}_0(\mathcal{H})\mathcal{P}_2(\mathcal{H})|K_0}=0.  \end{align}

Finally, we calculate the following entity
\begin{align}
     \braket{K_0|\mathcal{P}_1(\mathcal{H})\mathcal{P}_2(\mathcal{H})|K_0}=\frac{1}{\sqrt{N_1 N_2}}\Big[\braket{\Psi_1|\Psi_2}-c_1\braket{\Psi_1|\Psi_1}+c^2_0 c_1 -c_0 d_1\Big]
\end{align}
where we have used the fact that $a_0=\braket{K_0|\mathcal{H}|K_0}=\braket{K_0|\Psi_1}=c_0$.

Using the definitions \eqref{e2.22} and \eqref{e2.27}, one finds 
\begin{align}
    &\braket{\Psi_1|\Psi_1}=N_1+c_0^2\\
    &\braket{\Psi_1|\Psi_2}=c_1N_1+c_0 d_1
\end{align}
which finally yields the desired orthogonality
\begin{align}
\braket{K_0|\mathcal{P}_1(\mathcal{H})\mathcal{P}_2(\mathcal{H})|K_0}=0.
\end{align}

In a similar fashion, one can, in general, prove the orthogonality
\begin{align}
\braket{K_0|\mathcal{P}_m(\mathcal{H})\mathcal{P}_n(\mathcal{H})|K_0}=\delta_{mn}.
\end{align}
\section{Moments and return amplitude for Krylov operator growth}
\label{sec3}
The goal of this Section is to discuss the moments and the returns amplitude in the context of Krylov (operator) complexity \cite{Caputa:2025mii}. We set the stage by discussing the general framework of Krylov operator complexity in the BMN matrix model at strong mass deformation \cite{Roychowdhury:2026igc}. 
\subsection{Preliminaries}
The Krylov operator growth \cite{Parker:2018yvk} corresponds to the Heisenberg time evolution of an operator $\mathcal{O}(t)=e^{i\mathcal{H}t}\mathcal{O}(0)e^{-i \mathcal{H} t}$ in quantum mechanics. In the dual gravity picture, this would correspond to a massive probe particle following the geodesic in the bulk \cite{Caputa:2024sux}, \cite{Roychowdhury:2026sgg}, which corresponds to an insertion of a local operator $\mathcal{O}_0=\mathcal{O}(t=0)$ in the matrix model.

The formalism is rest on the map $\mathcal{O}_0 \rightarrow |\mathcal{O}_0)$, where $|\mathcal{O}_0)$ is a state in the operator Hilbert space. The other states $\mathcal{O}_n$ are constructed through successive application of the Liouvillian operator $\hat{\mathcal{L}}:=[\mathcal{H},]$, such that $\mathcal{O}_n = \hat{\mathcal{L}}^n\mathcal{O}_0$. This generates a time evolution
\begin{align}
\label{e3.1}
    \mathcal{O}(t)=\sum_{n=0}^\infty \frac{(it)^n}{n!}\hat{\mathcal{L}}^n\mathcal{O}_0=\sum_{n=0}^\infty \frac{(it)^n}{n!}\mathcal{O}_n
\end{align}
accompanied by a basis $\{ \mathcal{O}_n\}$, which has the elements;
\begin{align}
    \mathcal{O}_1=\hat{\mathcal{L}}\mathcal{O}_0 = [\mathcal{H},\mathcal{O}_0]~;~ \mathcal{O}_2=\hat{\mathcal{L}}^2\mathcal{O}_0 = [\mathcal{H},\mathcal{O}_1]~;~\cdots.
\end{align}

As mentioned before, the initial state corresponds to a pair of harmonic oscillator vacuum around their respective ground states \cite{Amore:2024ihm}-\cite{Huh:2024ytz}. This motivates us to choose 
\begin{align}
\label{e3.3}
    \mathcal{O}_0=\sqrt{\frac{2 \mu}{\pi}}e^{-\mu(\hat{x}^2+\hat{y}^2)}
\end{align}
such that, given a reference state $\ket{x,y}$ in the position basis, one finds
\begin{align}
\label{e3.4}
    \mathcal{O}_0 \rightarrow |\mathcal{O}_0)= \mathcal{O}_0 \ket{x,y}=\Psi_0(x,y)\ket{x,y}
\end{align}
with $\Psi_0(x,y)$ given in \eqref{e2.19}.

The initial operator \eqref{e3.3} is chosen to be a Gaussian one \cite{Hashimoto:2023swv} that can be justified under several grounds. At strong coupling $\mu \rightarrow \infty$, the eigenvalue distribution remains effectively localized near the origin ($x\sim 0, y \sim 0$), which causes the effective dynamics to map onto a system of localized harmonic oscillators. The ground state of a harmonic oscillator is a Gaussian state. The Gaussian operator \eqref{e3.3} (or the wave-packet/state) maximizes the overlap with the true ground state of the localized potential \eqref{e2.17}. 

With the above initial state \eqref{e3.4} constructed, one can compute a first few basis elements corresponding to $n=1,2$, which are given by
\begin{align}
    &\mathcal{O}_1 = \frac{2i \sqrt{2}\mu^{3/2}}{\sqrt{\pi}}(\hat{p}_x \hat{x}+\hat{p}_y \hat{y})e^{-\mu (\hat{x}^2+\hat{y}^2)}\\
  &\mathcal{O}_2=  \frac{2 \sqrt{2}\mu^{3/2}}{\sqrt{\pi}}\Big[\hat{p}^2_x+\hat{p}^2_y-2 \mu (\hat{p}_x \hat{x}+\hat{p}_y \hat{y})^2\Big]e^{-\mu (\hat{x}^2+\hat{y}^2)}-\frac{2 \sqrt{2}\mu^{7/2}}{\sqrt{\pi}}(k\hat{x}^2+\hat{y}^2)e^{-\mu (\hat{x}^2+\hat{y}^2)}\nonumber\\
  &-\frac{4 \sqrt{2}\mu^{3/2}}{\sqrt{\pi}}(\hat{x}^4+\hat{y}^4)e^{-\mu (\hat{x}^2+\hat{y}^2)}- \frac{8\sqrt{2}l\mu^{3/2}}{\sqrt{\pi}}\hat{x}^2 \hat{y}^2e^{-\mu (\hat{x}^2+\hat{y}^2)}\nonumber\\
  &+\frac{6\sqrt{2}\mu^{5/2}}{\sqrt{\pi}}\hat{y}^3e^{-\mu (\hat{x}^2+\hat{y}^2)}+\frac{6\sqrt{2}n\mu^{5/2}}{\sqrt{\pi}}\hat{x}^2 \hat{y} e^{-\mu (\hat{x}^2+\hat{y}^2)}
\end{align}
which is subject to the fact $[\hat{x},\hat{p}_x]=i=[\hat{y},\hat{p}_y]$ together with $\hat{p}_i=-i \partial_i$ and $\hbar=1$. Notice that here Hamiltonian $\mathcal{H}$ is an operator corresponding to \eqref{e2.5}.

Before we proceed further, it is customary to introduce the notion of the inner product between states in the operator Hilbert space
\begin{align}
\label{e3.7}
    (\mathcal{O}_n|\mathcal{O}_m)=\text{Tr}(\mathcal{O}^\dagger_n \mathcal{O}_m)=\int dx dy \bra{x,y}\mathcal{O}^\dagger_n \mathcal{O}_m \ket{x,y}
\end{align}
which clearly indicates that the initial state is normalized, that is, $(\mathcal{O}_0|\mathcal{O}_0)=1$.\\\\
\uline{\textbf{Constructing the Krylov basis:}}\\\\
The Krylov operator basis $\{ |K_n)\}$ is constructed following similar steps as mentioned before. Given an operator \eqref{e3.1}, one can expand it in the Krylov basis as \cite{Parker:2018yvk}
\begin{align}
\label{e3.8}
    \mathcal{O}(t)=\sum_n i^n \varphi_n(t)K_n
\end{align}
where $\{ |K_n)\}$ form an orthonormal basis, that is, $(K_m|K_n)=\delta_{mn}$.

Like in the case of the spread complexity, one has a corresponding Krylov chain criteria satisfied by the Krylov basis elements \cite{Parker:2018yvk}
\begin{align}
\label{e3.9}
    &A_{n+1}=\hat{\mathcal{L}}K_n-b_{n}K_{n-1}\\
    &A_{n+1}=b_{n+1}K_{n+1}.
\end{align}

By definition \eqref{e3.9}, it is clear that the diagonal entries are all zero
\begin{align}
    a_n=L_{nn}=(K_n|\hat{\mathcal{L}}|K_n)=b_{n+1}(K_n|K_{n+1})+b_n(K_n|K_{n-1})=0.
\end{align}

On the other hand, the off-diagonal elements produce non zero Lanczos coefficients
\begin{align}
    b_n=L_{n n-1} = (K_n| \hat{\mathcal{L}}|K_{n-1}).
\end{align}

Using \eqref{e3.8} in the Heisenberg time evolution of operators and using the Krylov chain criteria \eqref{e3.9}, one finally arrives at an equation for the coefficients $\varphi_n(t)$
\begin{align}
\label{e3.13}
    \partial_t \varphi_n (t)=b_n \varphi_{n-1}(t) -b_{n+1}\varphi_{n+1}(t).
\end{align}

The corresponding Krylov operator complexity is given by 
\begin{align}
\label{e3.14}
    \mathcal{K}_{\mathcal{O}}(t)=\sum_{n}n |\varphi_n(t)|^2.
\end{align}

In order to construct Krylov basis for operator growth, we follow the Gram-Schmidt procedure as before. We propose an expansion of the following form
\begin{align}
K_{n+1}=\mathcal{O}_{n+1}-c_{n}K_{n}-d_n K_{n-1}
\end{align}
which is subject to the fact that $K_0=\mathcal{O}_0$ and $K_{-1}=0$.

Setting $n=1$, we have the first Krylov element
\begin{align}
\label{e3.16}
    K_1=\frac{1}{\sqrt{N_1}}\Big[ \mathcal{O}_1-c_0K_0\Big]
\end{align}
where we calculate the coefficient $c_0$, which turns out to be
\begin{align}
\label{e3.17}
    c_0=(K_0|\mathcal{O}_1) =2\mu.
\end{align}

The associated normalization constant can be expressed as
\begin{align}
\label{e3.18}
    N_1=(K_1|K_1)=4\mu^2.
\end{align}

On a similar note, the next Krylov basis element can be expressed as
\begin{align}
\label{e3.19}
    K_2=\frac{1}{\sqrt{N_2}}\Big[ \mathcal{O}_2-c_1K_1-d_1 K_0\Big].
\end{align}

It is interesting to notice that the Krylov basis elements start seeing the different Fuzzy sphere vacuum starting at second order. This is perfectly reflected in the coefficients above, which are parametrized by $k$ and $l$, for example,
\begin{align}
\label{e3.20}
    & d_1=(K_0|\mathcal{O}_2)=-\frac{1}{2 \mu }\Big[ (k-7) \mu ^3+l+3\Big]\\
    &c_1=\frac{1}{4\mu^2}\Big[(\mathcal{O}_1|\mathcal{O}_2)-c_0 d_1 \Big]=\frac{1}{4\mu^2}\Big[ (k+25) \mu ^3+2 l+6\Big].
    \label{e3.21}
\end{align}

The normalization constant, on the other hand, can be expressed as
\begin{align}
    N_2=\frac{\mu ^4}{4} ((k-2) k+257) +\frac{\mu }{16}  (-128 l+27 n (n+2)-249)+\frac{(l^2+3)}{\mu ^2}.
\end{align}

With the above machinery at hand, we are now in a position to calculate the moments and the return amplitude for the matrix model at large mass deformation $\mu$. 
\subsection{Return amplitude and Lanczos coefficients}
In order to compute the return amplitude and moments, we notice the following expansions
\begin{align}
\label{e3.23}
    &\mathcal{O}(t)=\mathcal{O}_0+it \mathcal{O}_1-\frac{t^2}{2}\mathcal{O}_2+\cdots\\
    &\mathcal{O}(t)=\varphi_0 (t) K_0+i\varphi_1 (t)K_1-\varphi_2 (t)K_2+\cdots
    \label{e3.24}
\end{align}
that readily follow from \eqref{e3.1} and \eqref{e3.8} respectively.

The return amplitude \cite{Caputa:2025mii} for the Krylov operator growth is given by 
\begin{align}
    \mathcal{R}(t)=(\mathcal{O}(t)|\mathcal{O}_0).
\end{align}

With the choice of the initial state $|\mathcal{O}_0)=|K_0)$ and using the orthonormal property of the Krylov basis, it is straightforward to show that
\begin{align}
\label{e3.26}
    \mathcal{R}(t)=\varphi^\ast_0(t).
\end{align}

Using \eqref{e3.26}, one finally introduces the moments as follows
\begin{align}
\label{e3.27}
    \mathcal{M}_n=(-i)^n\frac{d^n}{dt^n}\mathcal{R}(t)\Big|_{t=0}.
\end{align}

Our goal is to calculate the function $\varphi_0(t)$, that satisfies one of the equations obtained by setting $n=0$ in \eqref{e3.13}
\begin{align}
\label{e3.28}
    &\partial_t \varphi_0 (t)= -b_1 \varphi_1 (t)
\end{align}
where $b_1(\mu)$ is the first non-zero Lanczos coefficient and is given by
\begin{align}
    b_1(\mu)=L_{10}=(K_1|\hat{\mathcal{L}}|K_0)=(K_1|[\mathcal{H},K_0])=(K_1|\mathcal{O}_1).
\end{align}

After some further simplification, we obtain
\begin{align}
\label{e3.30}
    b_1(\mu)=\frac{1}{\sqrt{N}_1}\Big[(\mathcal{O}_1|\mathcal{O}_1)-2\mu (K_0|\mathcal{O}_1)\Big]=2\mu.
\end{align}
As we show shortly in \eqref{e3.34}, this is precisely given by the second moment $\mathcal{M}_2(\mu)=b_1^2(u)$ \cite{Caputa:2025mii}, which follows from the return amplitude \eqref{e3.27}.

Using \eqref{e3.23} and \eqref{e3.24} and taking the inner product with $|K_1)$ on both sides, we find
\begin{align}
    \varphi_1(t)=2\mu t+\frac{it^2}{4 \mu}\Big[(\mathcal{O}_1|\mathcal{O}_2)-2\mu (K_0|\mathcal{O}_2) \Big]
\end{align}
which, after further simplification yields
\begin{align}
\label{e3.32}
    \varphi_1(t)=2 \mu  t+\frac{i t^2 }{4 \mu }\left((k+25) \mu ^3+2 l+6\right)+\mathcal{O}(\mu^3 t^3).
\end{align}

Using \eqref{e3.32} in \eqref{e3.28}, we finally obtain
\begin{align}
\label{e3.33}
    \varphi_0(t)&=1-2 \mu ^2 t^2-\frac{i t^3}{6}  \left((k+25) \mu ^3+2 l+6\right)+\mathcal{O}(\mu^4 t^4).
\end{align}

Notice that in the above solutions $\mu t$ is kept fixed in the limit $\mu \rightarrow \infty$ at $t\sim 0$. In particular $\varphi_0(0)=1$, which is consistent with the normalization of the state $|\mathcal{O}(t))$ at initial time $t=0$. Using above solutions \eqref{e3.32} and \eqref{e3.33}, together with 
\begin{align}
    \partial_t \varphi_1 =b_1 \varphi_0 (t) -b_2 \varphi_2 (t)
\end{align}
one finds the wave function corresponding to $n=2$ lattice site as
\begin{align}
\label{ee3.35}
    &\varphi_2 (t)=\kappa  \left[1-2 \mu ^2 t^2-\frac{i t^3}{6}  \left((k+25) \mu ^3+2 l+6\right)\right]-\frac{1}{b_2}\Big[ 2 \mu +\frac{i t \left((k+25) \mu ^3+2 l+6\right)}{2 \mu }\Big]\nonumber\\
    & \kappa = \frac{b_1}{b_2}=\frac{4\sqrt{k(k-2)+257 }}{k (k+20)+341}.
\end{align}

Considering a large $\mu$ expansion, this further reveals
\begin{align}
\label{ee3.36}
 \varphi_2 (t)=  -\Big[\frac{i (k+25) \sqrt{(k-2) k+257} }{k (k+20)+341} \Big]\mu  t -2\kappa  \mu ^2  t^2-\frac{i \kappa }{6}  (k+25) \mu ^3 t^3+\mathcal{O}(\mu^4 t^4).
\end{align}

\paragraph{An important note.} Looking at the solutions \eqref{e3.32}, \eqref{e3.33} and \eqref{ee3.35} (or \eqref{ee3.36}) in the limit $\mu \rightarrow \infty$, while keeping $\mu t$ finite and smaller than unity, it can be argued that the generic expression of the wave functions $\varphi_n(t)$ allows a \emph{polynomial} expansion in powers of the combination $\mu t$ where all the sub-leading terms are smaller as compared to their leading term in the expansion. This implies that, in the large mass deformation the wavefunctions are generically confined to a small time domain $t<1$ depending on the mass parameter $\mu$, which further implies a finite a size of the Krylov subspace. This observation is going to play significant role in the computation of the spectral function \eqref{ee4.37}.

It is clear by definition that all the odd moments $\mathcal{M}_1,\mathcal{M}_3,\cdots$ are zero by construction, which is consistent with the fact that one is looking at time evolution of an Hermitian operator $\mathcal{O}(t)$ \cite{Caputa:2025mii}. In other words, only \emph{even} moments are non vanishing.

The first non-zero (even) moment is given by
\begin{align}
\label{e3.34}
    \mathcal{M}_2 (\mu) = -\frac{d^2}{dt^2}\varphi^\ast_0(t)\Big|_{t=0}=4\mu^2=\alpha_{(2)} \mu^2.
\end{align}
Notice that the moment \eqref{e3.34} does not depend of the choice $k$ of the Fuzzy sphere vacuum. In other words, it is same for both $N=2$ and $N=4$ representations of the matrix model. Also notice that \eqref{e3.34} is an \emph{exact} expression since all higher order contributions as in \eqref{e3.33} would result in a vanishing contribution at $t=0$.

The next non-zero moment is given by \cite{Caputa:2025mii}
\begin{align}
\label{eee3.35}
    \mathcal{M}_4 (\mu)=b_1^4(\mu)+b_1^2(\mu)b_2^2(\mu).
\end{align}

The second non-zero Lanczos coefficient can be expressed as
\begin{align}
\label{e3.36}
    b_2(\mu)=(K_2|\hat{\mathcal{L}}|K_1)=(K_2|[\mathcal{H},K_1])=\frac{1}{2\mu}(K_2|\mathcal{O}_2)-(K_2|\mathcal{O}_1).
\end{align}

After evaluating the above entities in \eqref{e3.36}, one finds at LO in large $\mu$
\begin{align}
    b_2(\mu)=\frac{(k (k+20)+341) \mu }{2 \sqrt{(k-2) k+257 }}+\mathcal{O}(\mu^0).
\end{align}

This finally yields the moment 
\begin{align}
    &\mathcal{M}_4 (\mu)=\alpha_{(4)}\mu^4\\
    &\alpha_{(4)}=16 \Big[1+ \frac{(k (k+20)+341)^2 }{16((k-2) k+257)}\Big].
\end{align}

Clearly, the coefficient $\alpha_{(4)}$ depends on the representation of the Fuzzy sphere vacuum which is characterized by the parameter $k$. This is similar as found in the case of the spread complexity above. Below, we summarize them for different choices of the vacuum
\begin{equation}
\label{e3.40}
 \mathcal{M}_4(\mu)=
    \begin{cases}
      482.784\mu^4 & \text{PFS}~ (k=1/4)\\
      527.891\mu^4 & \text{IFS}~(k=1).
    \end{cases}       
\end{equation}
Notice that \eqref{e3.40} appears to be an exact expression in the sense of the expansion \eqref{e3.33} mentioned above, where the coefficient $\mu^4$ will only survive after taking four derivatives on $\varphi_0^\ast$, had we obtained the solution \eqref{e3.33} up to $\mathcal{O}(\mu^4 t^4)$, which in the present case is done, however, using a different combination \eqref{eee3.35}.

The above procedure could be extended at higher order in the time evolution which will generate (even) moments $\mathcal{M}_n(\mu)$ at higher order. Clearly, we see a general structure \eqref{e2.45} as in the case of the spread complexity of states, where the coefficients $\alpha_{(n)}$ (for $n$ even) are fixed by the Lanczos coefficients $b_n(\mu)$ that depend on the choice $k$ of the Fuzzy sphere vacuum and/or the representation of the $SU(2)$ algebra.

Finally, the Krylov operator complexity is given by
\begin{align}
\label{e3.43}
    \mathcal{K}_{\mathcal{O}}(t)=\sum_{n}n |\varphi_n(t)|^2=\zeta(k) \mu^2 t^2+\mathcal{O}(\mu^4 t^4).
\end{align}

Clearly, the operator complexity exhibits a quadratic growth at early time whose coefficient is characterized by the choice of the Fuzzy sphere vacuum
\begin{align}
    \zeta(k)=2+\frac{k(k-2)  (k+25)^2}{(k (k+20)+341)^2}+\frac{257 (k+25)^2}{(k (k+20)+341)^2}. 
\end{align}

The coefficient $\zeta(k)$ is precisely fixed by choosing the $k$ parameter which corresponds to different Fuzzy sphere vacuum, for example,
\begin{equation}
\label{e3.45}
  \zeta(k)=
    \begin{cases}
      3.36 & \text{PFS}~ (k=1/4)\\
      3.32 & \text{IFS}~(k=1).
    \end{cases}       
\end{equation}
These numbers are very close, therefore the early time growth in the matrix model does not have a qualitative difference for different representations of the Fuzzy sphere vacuum. 

\subsection{Remarks on orthogonal polynomials}
Orthogonal polynomials are defined in a spirit similar to the spread complexity grwoth. For the Krylov operator growth they are defined as \cite{Muck:2022xfc}
\begin{align}
    |K_n)=|\mathcal{P}_n(\hat{\mathcal{L}})K_0).
\end{align}
which satisfy an identity that follows from \eqref{e3.9}
\begin{align}
\hat{\mathcal{L}}\mathcal{P}_n(\hat{\mathcal{L}})=b_{n+1}\mathcal{P}_{n+1}(\hat{\mathcal{L}})+b_n\mathcal{P}_{n-1}(\hat{\mathcal{L}}).
\end{align}

To begin with, the obvious thing to notice is that $\mathcal{P}_0(\hat{\mathcal{L}})=1$ which maps the initial state $|K_0)$ to itself. The polynomial at the next order we propose is given by
\begin{align}
\label{e3.47}
    \mathcal{P}^{(k)}_1(\hat{\mathcal{L}})=\frac{1}{\sqrt{N_1}}(\hat{\mathcal{L}}^{(k)}-\gamma_0)
\end{align}
where $k$ corresponds to different representations of the Fuzzy sphere vacuum which would correspond to a different Liouvillian operator $\hat{\mathcal{L}}$.

Upon acting on the state $|K_0)$ and comparing with \eqref{e3.16}, we find the following
\begin{align}
\label{e3.48}
    \mathcal{P}^{(k)}_1(\hat{\mathcal{L}})=\frac{1}{2\mu}(\hat{\mathcal{L}}^{(k)}-2 \mu)~;~\gamma_0=c_0.
\end{align}

As a trivial check, one can show that (for a given representation) the orthogonality of the polynomials is satisfied by construction \eqref{e3.17}, that is,
\begin{align}
\label{e3.49}
   ( K_0| \mathcal{P}_0(\hat{\mathcal{L}}) \mathcal{P}_1(\hat{\mathcal{L}})|K_0)=\frac{1}{2\mu}\Big[(K_0|\hat{\mathcal{L}}K_0)-2\mu \Big]=\frac{1}{2\mu}\Big[(K_0|\mathcal{O}_1)-2\mu \Big]=0.
\end{align}

On a similar note, we propose the polynomial at next order is
\begin{align}
\label{e3.50}
    \mathcal{P}^{(k)}_2(\hat{\mathcal{L}})=\frac{1}{\sqrt{N_2}}\Big[(\hat{\mathcal{L}}^{(k)}-\gamma_1)(\hat{\mathcal{L}}^{(k)}-\gamma_2)+\gamma^2_2\Big].
\end{align}

Upon acting on the state $|K_0)$ (for a fixed $k$), we find the following
\begin{align}
    |K_2)=|\mathcal{P}_2(\hat{\mathcal{L}})K_0)=\frac{1}{\sqrt{N_2}}\Bigg[|\mathcal{O}_2)-(\gamma_1+\gamma_2)|\mathcal{O}_1)+\gamma_2 (\gamma_1+\gamma_2)|K_0) \Bigg].
\end{align}

Comparing with \eqref{e3.19}, one finds a relation identical to \eqref{e2.57}. Clearly, at LO in the large $\mu$ expansion, one trivially satisfies $\gamma_1+\gamma_2=c_1$ as in the case of the spread complexity of states, except for the fact that $c_1$ is a different value \eqref{e3.21} for the operator growth. This finally yields the following expression for the second order polynomial in large $\mu$ limit
\begin{align}
   &\mathcal{P}^{(k)}_2(\hat{\mathcal{L}})= \frac{1}{\sqrt{N_2}}\Bigg[\hat{\mathcal{L}}^{2(k)}_{PFS}+\beta_1(k) \mu\hat{\mathcal{L}}^{(k)}_{PFS}+ \beta_2(k)\mu^2\Bigg]\\
   &\beta_1(k)=-\frac{1}{4}(k+25)~;~\beta_2(k)=k+9.
\end{align}

Below, we summarize the second order polynomial for different representations of the Fuzzy sphere vacuum that corresponds to different $k$ values
\begin{equation}
  \mathcal{P}^{(k)}_2(\hat{\mathcal{L}})=
    \begin{cases}
     \frac{0.12}{ \mu ^2}\Bigg[\hat{\mathcal{L}}^2_{PFS}-6.31 \mu\hat{\mathcal{L}}_{PFS}+9.25 \mu^2\Bigg] & \text{PFS}~ (k=1/4)\\
    \frac{1}{8 \mu ^2}\Bigg[\hat{\mathcal{L}}^2_{IFS}-6.5 \mu\hat{\mathcal{L}}_{IFS}+10 \mu^2\Bigg] & \text{IFS}~(k=1).
    \end{cases}       
\end{equation}

Here, the subscripts refer to the Liouvillian operator $\hat{\mathcal{L}}$ in different representations of the matrix model that are characterized by different $k$ values. Notice that the pre-factors as well as $\gamma_{1,2}$ are different numbers and they correspond to different Fuzzy sphere vacuum.
The above procedure can be repeated $n$ times which yields an expression
\begin{align}
    \mathcal{P}^{(k)}_n(\hat{\mathcal{L}})\Big|_{\mu \gg 1}=\frac{d^{(k)}}{\mu^n}\Big[\hat{\mathcal{L}}^{n(k)} +\xi^{(k)}_{n-1}\mu\hat{\mathcal{L}}^{n-1(k)}+\xi^{(k)}_{n-2}\mu^2\hat{\mathcal{L}}^{n-2(k)} +\cdots +\xi^{(k)}_{0} \mu^{n}\Big]
\end{align}
which is structurally similar to the polynomial expression \eqref{e2.60}.\\\\
\uline{\textbf{Orthogonality of the polynomials:}}\\\\ 
Orthogonality of the polynomials follows in an identical fashion as that of the spread complexity of states discussed in the previous section. To begin with, let us recollect our results $\mathcal{P}_0(\hat{\mathcal{L}})=1$ along with \eqref{e3.49}. Next, we calculate the inner product
\begin{align}
     ( K_0| \mathcal{P}_0(\hat{\mathcal{L}}) \mathcal{P}_2(\hat{\mathcal{L}})|K_0)=\frac{1}{\sqrt{N_2}}\Bigg[(K_0|\mathcal{O}_2)-c_1(K_0|\mathcal{O}_1)+c_0 c_1-d_1 \Bigg]
\end{align}
which by means of \eqref{e3.17} and \eqref{e3.20} yields
\begin{align}
   ( K_0| \mathcal{P}_0(\hat{\mathcal{L}}) \mathcal{P}_2(\hat{\mathcal{L}})|K_0)=0. 
\end{align}

On a similar note, we find the following inner product
\begin{align}
    ( K_0| \mathcal{P}_1(\hat{\mathcal{L}}) \mathcal{P}_2(\hat{\mathcal{L}})|K_0)=\frac{1}{\sqrt{N_1 N_2}}\Big[(\mathcal{O}_1|\mathcal{O}_2)-c_1(\mathcal{O}_1|\mathcal{O}_1)+c^2_0 c_1-c_0 d_1 \Big].
\end{align}

From \eqref{e3.21} we find the following
\begin{align}
\label{e3.60}
 (\mathcal{O}_1|\mathcal{O}_2)=N_1 c_1+c_0 d_1.   
\end{align}

On the other hand, using \eqref{e3.17}, \eqref{e3.18} and \eqref{e3.30}, we find the following
\begin{align}
\label{e3.61}
    (\mathcal{O}_1|\mathcal{O}_1)=N_1+c_0^2.
\end{align}

Using above relations \eqref{e3.60} and \eqref{e3.61}, one finds
\begin{align}
    ( K_0| \mathcal{P}_1(\hat{\mathcal{L}}) \mathcal{P}_2(\hat{\mathcal{L}})|K_0)=0.
\end{align}

One can construct higher order polynomials to show orthogonality in general as
\begin{align}
    ( K_0| \mathcal{P}_m(\hat{\mathcal{L}}) \mathcal{P}_n(\hat{\mathcal{L}})|K_0)=\delta_{mn}.
\end{align}

Before we conclude our discussion on orthogonal polynomials, it is worth emphasizing the fact that the orthogonality of the polynomials should hold irrespective of the large $\mu$ expansion. In other words, both for the spread complexity of states as well as the Krylov operator growth, the proof of the orthogonality of the polynomials does not require a large $\mu$ limit and should follow by construction for all values of the deformation parameter $\mu$.
\section{Spectral function and density of states}
\label{sec4}
The purpose of this section is to build up an algorithm in the context of Krylov operator growth that would eventually lead towards spectral function and density of states. This is an algorithm parallel to the Lanczos algorithm, which is, however, richer in its content and contains several recursion relation of entities like functions of the second kind and so on. The algorithm is developed using a number of steps. In the following, we will enumerate them explicitly for the BMN matrix model at large mass deformation. All these will be used finally to compute the spectral function and the density of states in the matrix model.
\subsection{Building blocks of the many body dynamics}
The key elements of the Krylov complexity are the wavefunctions $\varphi_n(t)$ that satisfy the Schrodinger equation \eqref{e3.13}. An alternative way to solve these wavefunctions without going into iteration is to introduce the notion of the truncated Laplace transform 
\begin{align}
\label{e3.64}
    \mathfrak{c}_n(z)=\int_0^T dt \varphi_n(t)e^{-zt}~;~z=\omega+i \epsilon
\end{align}
where the integral is performed over an interval $[0,T]$.

To understand why the primary contribution to the truncated Laplace transform \eqref{e3.64} is localized near $t \sim 0$, consider the interplay between the frequency parameter $z = \omega + i\epsilon$ and the intrinsic energy scale $\mu$, directly building upon the dimensionless parameter $(\mu t)$ discussed in Section \ref{sec2.2}.
For the early-time region $t \ll 1/\mu$ (where $\mu t \ll 1$), the wavefunctions behave as $\varphi_n(t) \sim (\mu t)^n$, leading to an explicit power-law integrand $t^n e^{-z t}$. Conversely, for intermediate and late times $t \sim \mathcal{O}(1/\mu)$ or $t \gg 1/\mu$, the mass parameter acts as a heavy frequency cutoff ($\mu \gg 1$), causing the operator wavefunctions $\varphi_n(t)$ to rapidly oscillate and decay due to destructive phase interference across the Krylov chain. 
When evaluating the integral over $[0, T]$, the lower boundary $t \to 0$ yields a dominant boundary contribution proportional to $1/z^{n+1}$. In contrast, contributions from $t \sim T$ are exponentially suppressed by $e^{-zT}$ for $\text{Re}(z)>0$. Furthermore, in the complex frequency domain where $|z| \sim \mu$, the weight function $e^{-z t}$ decays on a timescale $\tau \sim 1/|z| \sim 1/\mu$, restricting the non-negligible support of the integrand precisely to the boundary regime $t \lesssim 1/\mu \ll 1$. Thus, the small-$t$ approximation is self-consistent across all frequency domains $\omega$, with the large-$\mu$ bound guaranteeing that boundary dynamics at $t \approx 0$ dictate the structural poles and divergence properties in $\omega$-space.

Substituting \eqref{e3.64} into \eqref{e3.13}, one finds the following relation \cite{Muck:2022xfc}
\begin{align}
\label{e3.65}
    z\mathfrak{c}_n(z)=b_n \mathfrak{c}_{n-1}(z)-b_{n+1}\mathfrak{c}_{n+1}(z)+\delta_{n0}
\end{align}
which is identical to the Krylov chain criterion and can be solved by the method of recurrence. Below we evaluate these functions directly by substituting the wavefunction.\\\\
\uline{$\bullet$ $n=0$ case:}
The corresponding Laplace transform is given by
\begin{align} 
\label{e3.66}
    \mathfrak{c}_0(z)=\int_0^\infty dt \varphi_0(t)e^{-zt}.
\end{align}

Using \eqref{e3.33} and after evaluating the integral \eqref{e3.66}, we find
\begin{align}
\label{e3.67}
    \mathfrak{c}_0(z)=-\frac{i\mu ^3}{z^4}(k+25)-\frac{4 \mu ^2}{z^3}+\frac{z^3-2 i (l+3)}{z^4}.
\end{align}\\
\uline{$\bullet$ $n=1$ case:} The coefficient of the next order can be obtained as
\begin{align}
     \mathfrak{c}_1(z)=\int_0^\infty dt \varphi_1(t)e^{-zt}
\end{align}
which by means of \eqref{e3.32} yields the following
\begin{align}
\label{e3.70}
    \mathfrak{c}_1(z)=\frac{1}{2 \mu  z^3}\Big[ i (k+25) \mu ^3+2 i (l+3)+4 \mu ^2 z\Big].
\end{align}

After performing a large $\mu$ expansion, this further yields
\begin{align}
\label{e3.71}
  \mathfrak{c}_1(z)=  \frac{i  \mu ^2}{2 z^3}(k+25)+\frac{2 \mu }{z^2}.
\end{align}

Using \eqref{e3.30}, \eqref{e3.67} and \eqref{e3.70}, it is straightforward to show
\begin{align}
    z \mathfrak{c}_0(z)=-b_1\mathfrak{c}_1(z)+1
\end{align}
which satisfies the identity \eqref{e3.65} for $n=0$.\\\\
\uline{$\bullet$ $n=2$ case:} On a similar note, using \eqref{ee3.35} at next order we find the coefficient
\begin{align}
    \mathfrak{c}_2(z)&=-\frac{1 }{2 b_2 \mu z^4}\Big[2 \mu  z^3 (2 \mu -b_2 \kappa )+8 b_2 \kappa  \mu ^3 z+i z^2 \left((k+25) \mu ^3+2 l+6\right)\nonumber\\
    &+2 i b_2 \kappa  \mu  \left((k+25) \mu ^3+2 l+6\right)\Big].
\end{align}

Considering a large $\mu$ expansion, this further yields
\begin{align}
\label{e3.74}
    \mathfrak{c}_2(z)=-\frac{i b_1  \mu ^3}{b_2 z^4}(k+25)-\frac{i \mu^2}{2b_2 z^2}(k+25)-\frac{4 b_1 \mu^2}{b_2 z^3}.
\end{align}

Using \eqref{e3.67} and \eqref{e3.71}, the above relation \eqref{e3.74} reveals the identity \eqref{e3.65}
\begin{align}
\label{e3.75}
    b_2 \mathfrak{c}_2(z)=b_1 \mathfrak{c}_0(z)-z\mathfrak{c}_1 (z).
\end{align}\\
\uline{$\bullet$ Large $\mu$ generalization:}
In order to find a generalized expression in the large $\mu$ limit, let us notice the following from \eqref{e3.71} and \eqref{e3.75}
\begin{align}
    &b_1 \mathfrak{c}_1(z)=\frac{i \mu^3}{z^3}(k+25)\\
    &b_2 \mathfrak{c}_2(z)=-\frac{2i \mu^4}{z^4}(k+25).
\end{align}

The above procedure can be extended for higher values of $n$. For example, using the recurrence relation \eqref{e3.65}, one finds the following relations
\begin{align}
    & b_3 \mathfrak{c}_3(z)=\frac{i \mu^3}{z^3}(k+25)\Big[\frac{(k (k+20)+341)}{4\sqrt{(k-2) k+257 }}-\frac{b_1}{b_2} \Big]\\
    &b_4 \mathfrak{c}_4(z)=-\frac{i \alpha b_1 \mu^4}{b_2 z^4}(k+25).
\end{align}

Given the expressions above in the large $\mu$ limit, we propose the following generalization (at LO) for the Laplace transform of the wave function $\varphi_n(t)$
\begin{align}
    &b_n \mathfrak{c}_n(z)=i\zeta_{n}(k+25)\frac{\mu^3}{z^3}~;~n=1,3,\cdots\\
    &b_n \mathfrak{c}_n(z)=-i\chi_{n}(k+25)\frac{\mu^4}{z^4}~;~n=2,4,\cdots.
\end{align}\\
\uline{$\bullet$ Memory function:} Next, we introduce the memory function $\mathfrak{R}_1(z)$ of the BMN matrix model along the lines of \cite{Muck:2022xfc}-\cite{Muck:2024fpb}. We define functions for a given $n$
\begin{align}
\label{e3.82}
    \mathfrak{R}_n(z)=\frac{b_n \mathfrak{c}_n(z)}{\mathfrak{c}_{n-1}(z)}~;~n>0.
\end{align}

Below, we list the functions \eqref{e3.82} for different choices of $n$, for example,
\begin{align}
    &\mathfrak{R}_1(z)|_{\mu \gg 1}=-z\\
    &\mathfrak{R}_2(z)|_{\mu \gg 1}=-\frac{b_1^2}{z}\\
    &\mathfrak{R}_3(z)|_{\mu \gg 1}=-\frac{b_2^2 z}{b^2_1}-z.
\end{align}
Notice that the odd functions are independent of the mass deformation parameter at LO, whereas, on the other hand, the even functions scale quadratically with $\mu$.\\\\
\uline{$\bullet$ Monic version:} There exists a parallel approach to recursion method in which the traditional Lanczos coefficients $b_n$ are replaced by monic coefficients $\Delta_n (n>0)$, which satisfy 
\begin{align}
\label{e3.85}
    K_{n+1}=\hat{\mathcal{L}}K_n-\Delta_{n}K_{n-1}.
\end{align}

The monic coefficients can be expressed in terms of functions \eqref{e3.82} defined above 
\begin{align}
\label{e3.87}
    \Delta_n = z \mathfrak{R}_n(z)+\mathfrak{R}_n(z)\mathfrak{R}_{n+1}(z).
\end{align}

Below we explicitly calculate \eqref{e3.87} for different $n$ in the large $\mu$ limit, for example,
\begin{align}
    &\Delta_1=b_1^2 \\
    &\Delta_2=b_2^2.
\end{align}

In summary, we find an expression that relates these different coefficients for large $\mu$
\begin{align}
    \Delta_n=b_n^2
\end{align}
which is in accordance with that proposed in \cite{Muck:2022xfc}.\\\\
\uline{$\bullet$ Functions of the second kind:} Next we introduce functions of the second kind \cite{Muck:2022xfc} for the matrix model and calculate them in the large $\mu$ limit. These functions are defined in terms of Laplace transform that are introduced in \eqref{e3.64}
\begin{align}
\label{e3.90}
    \mathcal{Q}_n(z)=\frac{\sqrt{h_n}}{i^{n+1}}\mathfrak{c}_n(-iz).
\end{align}

Here, $h_n=(K_n|K_n)$ (with $h_0=1$) and is not necessarily one for monic version ($n \geq 1$). These functions are related to monic coefficients \cite{Muck:2022xfc} as follows
\begin{align}
    \Delta_n=\frac{h_n}{h_{n-1}}=b^2_n.
\end{align}

The \emph{resolvent} is defined for $n=0$, which for the present case yields
\begin{align}
\label{e3.91}
    \mathcal{Q}_0(z)=-\frac{\mu^3}{z^4}(k+25)+\frac{4 \mu^2}{z^3}+\frac{-2 l+z^3-6}{z^4}.
\end{align}

For $n=1,2$, the other functions \eqref{e3.90} can be expressed as
\begin{align}
   &\mathcal{Q}_1(z)=-\frac{b_1\mu ^2}{2z^3}(k+25)+\frac{2 b_1\mu}{z^2} \\
   &\mathcal{Q}_2(z)=\frac{\mu ^3 b^2_1}{z^4}(k+25)+\frac{ib_1 \mu ^2 }{2  z^3}(8 i b_1 +i k z+25 i z)\nonumber\\&+\frac{2 b_1\mu }{z}+\frac{b_1^2(6  +2l- z^3)}{z^4}.
   \label{e3.93}
\end{align}

Combining \eqref{e3.91}-\eqref{e3.93}, one finds following recursion relations in the large $\mu$ limit
\begin{align}
\label{e3.95}
z\mathcal{Q}_0=\mathcal{Q}_1+1~;~z\mathcal{Q}_1=\mathcal{Q}_2+\Delta_1\mathcal{Q}_0.
\end{align}

The above relations \eqref{e3.95} can be generalized for arbitrary $n$, which yields 
\begin{align}
\label{e3.96}
    z\mathcal{Q}_n(z) = \mathcal{Q}_{n+1}(z)+\Delta_n \mathcal{Q}_{n-1}(z)+\delta_{n0}.
\end{align}

In summary, in the above exercise \eqref{e3.95}, we explicitly verify the recursion relation \cite{Muck:2022xfc} in the context of the BMN matrix model at large mass deformation. Although we have estimated functions of the second kind $\mathcal{Q}_n(z)$ for a few orders, however, the above formalism could in principle be extended to arbitrary order $n$. The entire exercise also gives us confidence that functions of the second kind \eqref{e3.90} are indeed evaluated correctly.

In the following, we will be primarily interested in the resolvent \eqref{e3.91} that is identified with the spectral function (and hence the spectral density) of the matrix model. Clearly, the resolvent \eqref{e3.91} is different for different representation of the Fuzzy sphere
\begin{equation}
\label{e3.92}
  \mathcal{Q}_0(z)=
    \begin{cases}
      -\frac{101 \mu ^3}{4 z^4}+\frac{4 \mu ^2}{z^3}+\frac{z^3-8}{z^4}+\cdots & \text{PFS}~ (k=1/4, l=1)\\
      -\frac{26 \mu ^3}{z^4}+\frac{4 \mu ^2}{z^3}+\frac{z^3-12}{z^4}+\cdots & \text{IFS}~(k=1,l=3).
    \end{cases}       
\end{equation}
\subsection{Spectral function and resonance}
The resolvent in Krylov complexity paves the way to study the spectral function and low energy dynamics in a quantum many body system. The resolvent operator \eqref{e3.91} defines the Green's function as well as it bridges the dynamical growth of operators to the spectral properties of the matrix model by introducing the spectral density function. 

Given the Liouvillian $\hat{\mathcal{L}}$, the resolvent \eqref{e3.91} can be expressed as
\begin{align}
\label{e4.35}
    \mathcal{Q}_0(z)=(\mathcal{O}|\frac{1}{z-\hat{\mathcal{L}}}|\mathcal{O}).
\end{align}

By taking the Laplace transform of the auto-correlation function in the time domain
\begin{align}
    G(t)=(\mathcal{O}(t)|\mathcal{O}(0))=(\mathcal{O}|e^{i\hat{\mathcal{L}t}}|\mathcal{O})
\end{align}
one can precisely arrive at the resolvent \eqref{e4.35}.

Given the state of the art, we introduce the density of states $\nu(\omega)$ which is defined in terms of the spectral function $\mathcal{A}(\omega)$ which is the imaginary part of the resolvent $\mathcal{Q}_0(z)$ \cite{Muck:2022xfc}
\begin{align}
\label{ee4.37}
  \mathcal{A}(\omega)=  \frac{d \nu(\omega)}{d\omega}=-\frac{1}{\pi}\lim_{\epsilon \rightarrow 0+}\Im [\mathcal{Q}_0 (\omega +i \epsilon)]=\frac{1}{\pi}\lim_{\epsilon \rightarrow 0+}\Re [c_0(\epsilon -i \omega)].
\end{align}

Before we proceed further, it is worth highlighting the fact that \eqref{ee4.37} is obtained using the initial wave function \eqref{e3.33}, which is truncated at $\mathcal{O}(\mu^4 t^4)$, where $\mu t <1$ in the limit $\mu \rightarrow \infty$ and $ t\sim 0$. In principle, the zeroth order wave function \eqref{e3.33} receives corrections in an expansion in time $t$ along the Krylov basis. However, in the strong coupling, all the higher order effects has to be smaller than its previous term \eqref{ee1.1} in order for the solution to make sense. In summary, the sub-leading corrections (at quartic order) are even smaller and can be ignored to some approximation, which finally results in \eqref{e4.37}. 

The density of states $\nu(\omega)$ is expressed in terms of the integral of the spectral function $\mathcal{A}(\omega)$ in the continuum limit, that is,
\begin{align}
\label{e4.38}
    \nu(\omega)=\int d\omega \mathcal{A}(\omega)=-\frac{1}{\pi}\lim_{\epsilon \rightarrow 0+}\Im \int d\omega  \mathcal{Q}_0 (\omega +i \epsilon).
\end{align}

Considering $\epsilon$ to be small enough ($\epsilon \mu \ll 1$), we find the spectral function
\begin{align}
\label{e4.37}
    \mathcal{A}^{(k,l)}(\omega)=-\frac{1}{\pi}\lim_{\epsilon \rightarrow 0+}\frac{\epsilon \Pi^{(k,l)}(\omega)}{\omega^2+\epsilon^2\Sigma^{(k,l)}(\omega)}.
\end{align}
The indices $k,l$ are characteristics of different representations of the Fuzzy sphere vacuum. 

The functions above can be expressed as
\begin{align}
\label{e4.40}
    &\Pi^{(k,l)}(\omega)=\frac{1}{\omega^3}\Big[ 4\mu ^3 (k+25) -12 \mu ^2 \omega +8 l-\omega ^3+24\Big]\equiv \frac{g_{kl}(\omega)}{\omega^3}\\
    &\Sigma^{(k,l)}(\omega)=\frac{20\mu ^3 (k+25)-40 \mu ^2 \omega +40 (l+3)-\omega ^3}{4\mu ^3 (k+25)-12 \mu ^2 \omega +8 l-\omega ^3+24}.
    \label{e4.41}
\end{align}

In the following, we explore different cases.\\\\
\uline{$\bullet$ Outside the resonance ($\omega \neq 0$):} To begin with, we consider the case away from the resonance. In this limit, kepping $\omega$ finite, the spectral function disappears trivially
\begin{align}
    \mathcal{A}^{(k,l)}(\omega)=-\frac{1}{\pi}\lim_{\epsilon \rightarrow 0+}\epsilon \frac{\Pi^{(k,l)}(\omega)}{\omega^2}=0.
\end{align}

Let us probe further into the behavior of the spectral function \eqref{e4.37} at large frequencies $\omega \rightarrow \infty $ while keeping the mass deformation $\mu$ finite and the broadening factor $\epsilon$ small. Under this assumption, one finds the following expression for the spectral function
\begin{align}
\label{e4.41}
     \mathcal{A}^{(k,l)}(\omega)|_{\omega \rightarrow \infty}=\frac{1}{\pi}\lim_{\epsilon \rightarrow 0+}\frac{\epsilon }{\omega^2+\epsilon^2}=\frac{\epsilon}{\pi \omega^2}.
\end{align}

The vanishing of the spectral function \eqref{e4.37} in the UV is incredibly important and reassuring result for the matrix model. For physical systems, the spectral function must vanish at high frequencies, reassuring the fact that the total integral (the sum rule) does not blow up at high energies. The $\frac{1}{\omega^2}$ behavior or the power law fall-off \eqref{e4.41} ensures that high energy nonphysical states do not contribute infinitely to the systems ground state properties. Such a power law behavior is known as the ``high frequency tail'' which unlike the low energy collective excitations represents the incoherent background. This tells us that if one pumps the system with a large amount of excitations $\omega \gg 1$, the probability of creating a stable resonant excitation drops off as a power law as depicted above \eqref{e4.41}. The energy is dissipated in the system as short lived continuum of multi-particle states.\\\\
\uline{$\bullet$ Near the resonance ($\omega \sim 0$):} This is the most subtle point which we should address carefully. We express \eqref{e4.37} in the following form that would be relevant in what follows
\begin{align}
\label{e4.44}
    \mathcal{A}^{(k,l)}(\omega)=-\frac{1}{\pi}\lim_{\epsilon \rightarrow 0+}\frac{\epsilon g_{kl}(\omega)}{\omega^3(\omega^2+\epsilon^2 \Sigma^{(k,l)}(\omega))}.
\end{align}

Let us take the limit $\omega \rightarrow 0$ for \emph{finite} broadening parameter $\epsilon >0$, which yields
\begin{align}
\label{e4.43}
    \mathcal{A}^{(k,l)}(\omega \sim 0)=\mathcal{A}_F^{(k,l)}(\omega \sim 0)+\mathcal{A}^{(k,l)}_Q(\omega \sim 0)+\mathcal{A}_P^{(k,l)}(\omega \sim 0).
\end{align}

Below, we elaborate each of these terms in detail. The first two expressions are evaluated first taking a $\omega \rightarrow 0$ limit and thereby taking $\epsilon \rightarrow 0+$ limit. The above procedure is useful to isolate the pole near $\omega=0$ for a finite but small broadening parameter $\epsilon$.\\\\
\uline{(i) Fatal divergence:} The fatal IR divergence corresponds to the first term in \eqref{e4.43}
\begin{align}
\label{e4.46}
 \mathcal{A}_F^{(k,l)}(\omega \sim 0)|_{\mu \gg 1}=-\frac{1}{\pi} \lim_{\epsilon \rightarrow 0+}\lim_{\omega \rightarrow 0} \frac{4\epsilon \mu^3 (k+25) }{\omega^3(\omega^2+5\epsilon^2 )} \approx -\lim_{\epsilon \rightarrow 0+}\frac{4 \mu^3}{5 \pi \epsilon \omega^3}(k+25) 
\end{align}
where $\epsilon \mu^3$ is finite in the limit of small $\epsilon$. The presence of such divergence indicates that the system has nonphysical infinite density of states at low energy that cannot be normalized.\\\\
\uline{(ii) Quadratic pole:} The quadratic pole appearing at next order can be expressed as
\begin{align}
\label{e4.47}
  \mathcal{A}^{(k,l)}_Q(\omega \sim 0)|_{\mu \gg 1}= \lim_{\epsilon \rightarrow 0+} \lim_{\omega \rightarrow 0}\frac{12 \epsilon \mu^2}{\pi \omega^2(\omega^2+5\epsilon^2)}=\lim_{\epsilon \rightarrow 0+} \frac{12 \mu^2}{5 \pi \epsilon \omega^2}+\mathcal{O}(\omega^2/\epsilon^2).
\end{align}
Like the previous divergence, the quadratic pole \eqref{e4.47} must also be tamed accordingly in order to preserve the sum rule or the finite total probability.\\\\
\uline{(iii) collective state:} Finally, we note down the contribution in the spectral function which allows physical states at low frequencies
\begin{align}
    &\mathcal{A}_P^{(k,l)}(\omega \sim 0)|_{\mu \gg 1}=\frac{1}{\pi}\lim_{\epsilon \rightarrow 0+}\frac{\epsilon}{\omega^2+\epsilon^2 \Sigma^{(k,l)}(0)}\\
    &\Sigma^{(k,l)}(0)=5.
\end{align}

Next, we take the limit $\epsilon \rightarrow 0+$, which by virtue of the properties of the delta function reveals the following expression for the spectral function
\begin{align}
\label{e4.50}
    \mathcal{A}_P^{(k,l)}(\omega \sim 0)|_{\mu \gg 1}=\frac{1}{\Sigma^{(k,l)}(0)}\lim_{\epsilon \rightarrow 0+}\frac{1}{\pi}\frac{\epsilon}{\Big( \frac{\omega}{\sqrt{\Sigma^{(k,l)}(0)}}\Big)^2+\epsilon^2}=\frac{1}{\sqrt{5}}\delta(\omega).
\end{align}

The pre-factor in \eqref{e4.50} corresponds to the weight or the quantum probability, which is $\frac{1}{\sqrt{5}}$ in the strict large $\mu$ limit. The presence of the delta function peak in the spectral function \eqref{e4.50} is indicative of a stable long lived ground state excitation in the matrix model. The pre-factor $\frac{1}{\sqrt{5}}$ (also known as the collective weight factor $Z$) is the measure of the how much of the bare particle's identity survives in the presence of strong interactions (characterized by the coupling $\mu$) of the dense surrounding quantum medium \cite{coleman_2015}-\cite{agd_1975}. Typically, for systems such as Fermi liquids, $0<Z<1$, which is precisely the case here.

Let us try to understand from the nature of the spectral function \eqref{e4.44}, why there could be stable collective peak in the theory and what are their consequences for a gaped system as in the example of the matrix model. Looking at \eqref{e4.44}, one can identify the term in the denominator $\epsilon^2 \omega^3 \Sigma^{(k,l)}(\omega)$ which can be identified with the decay rate or the imaginary part of the self energy in the quantum many body dynamics, that is, $\Sigma_I(\omega)\propto \omega^3$. Since the decay rate varies with higher power of the frequency, the probability that the excitation will undergo a scattering to other states vanishes rapidly as we move deep into the IR. The above is the consequence of the phase space protection, which states that at zero frequency the particle scatters with nothing and is therefore absolutely stable.

Notice that the matrix model we are describing here is QFT with a large mass gap. Similar systems exists in standard condensed matter setup where low energy collective excitations can exist in the presence of a large mass gap, for example, in topological insulators \cite{bernevig_2013} or certain class of superconductors \cite{tinkham_1996}-\cite{bruus_flensberg_2004} where the bulk of the material has a large mass gap while the edge or the boundary states contain gapless excitations. Because of the large mass gap, such collectives do not scatter and therefore do not decay which is perfectly compatible with the matrix model setup as alluded above.

It is interesting to note that although the collective weight factor $Z<1$ for the matrix model, the corresponding lifetime $\tau \propto 1/\omega^3$ scales differently from the conventional Fermi liquid, that is, $\tau \propto 1/\omega^2$. In other words, the effective damping rate scales even faster than the conventional Fermi liquid, and as a result the low energy (collective) excitations are more stable than those in a standard metal, indicating a non-Fermi liquid behavior. Comparing between the high and low frequency behavior of the spectral function \eqref{e4.44}, one can describe the collective excitations in the matrix model as a cross-over from a heavily damped state in the UV to the pristine well behaved quantum liquid in the deep IR. 
\subsection{A diagnostic of the IR divergences}
In a quantum many body system, IR divergences are indicative of the fact that in the calculation one typically ignores the effects due to finite life time of the collective modes near resonance. In other words, the bare or unregulated spectral function treats states near the resonance with infinite life time, ignores interactions, and does not properly account for decay channels. In standard treatment in quantum many body physics, the ill behaved spectral function is cured by a proper shift from the bare representation to the standard ``dressed representation''. This is achieved through proper (complex) self energy computation using standard techniques like many body perturbation theory, maximum entropy method etc. However, for the purpose of the present paper, we are not going to implement these methods for the matrix model and leave them for future investigation.

In what follows, we diagnose the pole structure of \eqref{e4.44} in more detail as we take the limit $\epsilon \rightarrow 0+$. This would further help us to gain insights about the collective modes at resonance. The spectral function \eqref{e4.44} can be schematically expressed as (we remove superscripts $(k,l)$ for simplicity)
\begin{align}
\label{e4.51}
    \mathcal{A}(\omega)=-\frac{1}{\pi} \lim_{\epsilon \rightarrow 0+}\frac{\epsilon(a+b \omega+c \omega^3)}{\omega^3(\omega^2+\epsilon^2 \Sigma(\omega))}
\end{align}
where the details of the coefficients $a$, $b$ and $c$ can be read off from \eqref{e4.40}.\\\\
\uline{$\bullet$ Isolating the resonance:} Let us first isolate the resonance peak at $\omega=0$, which is the conventional delta function peak. Using the properties of the delta function
\begin{align}
    \lim_{\epsilon \rightarrow 0+}\frac{1}{\pi}\frac{\epsilon}{x^2+\epsilon^2}=\delta(x)
\end{align}
we find the following expression for the entity
\begin{align}
\label{e4.53}
    \lim_{\epsilon \rightarrow 0+}\frac{1}{\pi}\frac{\epsilon}{\omega^2+\epsilon^2 \Sigma(\omega)}=\frac{1}{\sqrt{\Sigma(\omega)}}\delta(\omega).
\end{align}\\
\uline{$\bullet$ Evaluating the full spectral function:} Next, we evaluate the full spectral function 
\begin{align}
\label{e4.54}
    \mathcal{A}(\omega)=-\frac{(a+b \omega+c \omega^3)}{\omega^3}\times \frac{1}{\sqrt{\Sigma(\omega)}}\delta(\omega)=-f(\omega)\delta(\omega)
\end{align}
where we have considered the $\epsilon \rightarrow 0+$ limit by virtue of \eqref{e4.53}. 

Clearly, due to the presence of the delta function in \eqref{e4.54}, the function $f(\omega)$ has to be evaluated at $\omega =0$, which gives rise to fatal divergences due to $\frac{1}{\omega^3}$ term. In order to define density of states \eqref{e4.38}, one has to consider \eqref{e4.54} as a distribution under the integral and expand delta function into its derivatives using the integration by parts. The $\frac{1}{\omega^3}$ pole \eqref{e4.46} or $\frac{1}{\omega^2}$ pole structure \eqref{e4.47} are indicative of the fact that these collective modes do not represent standard propagating degrees of freedom of a quantum many body system rather they correspond to highly ``collective'' non-Fermi liquid states.

To understand the physical properties of these density low energy excitations, we explore the pole structure of the retarded Green's function or the resolvent $\mathcal{Q}_0(\omega)$. Clearly, we have a pole of order three at $\omega =0$ which reveals a non-propagating static structural feature at low frequency. The other is the $\epsilon$ dependent pole $\omega_{(\pm)} = \pm i \epsilon \sqrt{\Sigma(0)}\approx \pm i \epsilon\sqrt{5}$. Since these poles are purely imaginary, they do not represent any propagating degrees of freedom rather in the limit $\epsilon \rightarrow 0+$, the corresponding degrees of freedom (or collectives) become completely immobile or stationary. These are precisely the signature of the presence of the so called ``diffusion modes'' or ``relaxation modes'' with relaxation time $\tau \sim 1/\epsilon \rightarrow \infty$.\\\\
\uline{$\bullet$ Regularization:} We now comment on the possible regularization of the spectral function \eqref{e4.51}. In order to define a physical spectral function that satisfies the sum rule, that is, $\int \mathcal{A}(\omega)d\omega =1$, there are different regularization schemes that can be adopted.

In actual physical systems, $\epsilon$ is finite and non-zero, indicating a physical cut-off such as the scattering rate $\tau^{-1}$. By introducing $\Gamma(\omega)=\epsilon \sqrt{\Sigma(\omega)}$ as a frequency dependent damping parameter or memory function, one can re-express the spectral function
\begin{align}
\label{e4.55}
    \mathcal{A}(\omega)=-\frac{1}{\pi} \frac{\Gamma(\omega)(a+b \omega+c \omega^3)}{\omega^3\sqrt{\Sigma(\omega)}(\omega^2+\Gamma^2(\omega))}.
\end{align}
At low frequencies $\omega \ll \Gamma (\omega)$, the spectral function \eqref{e4.55} exhibits a power law divergence $\mathcal{A}(\omega)\sim \frac{a}{\omega^3 \Gamma (\omega)}$ that is typical of a fractional or non-Fermi liquid transport. 

Considering the dominant pole (of order $3$) contribution at low frequencies, the \emph{regularized} spectral function corresponding to \eqref{e4.54} can be expressed as
\begin{align}
    \mathcal{A}^{(reg)}(\omega \sim 0)/\sqrt{\Sigma(0)}=\mathcal{A}(\omega)+\mathcal{A}_{ct}
\end{align}
where we perform a rescaling by $\sqrt{\Sigma(0)}$ and the counter-term can be expressed as
\begin{align}
   \mathcal{A}_{ct}= \frac{4\mu^3}{\omega^3}(k+25)\delta(\omega)= \frac{4\mu^3}{\omega^3}(k+25)\delta(\omega).
\end{align}

This finally yields the physical density of states (D.O.S.) with $c=-1$ (see \eqref{e4.40})
\begin{align}
\label{e4.58}
    \text{D.O.S.}=\int d\omega \mathcal{A}^{(reg)}(\omega)=1.
\end{align}
It is interesting to notice that, unlike previous observables computed for the matrix model, the D.O.S. does not depend on the representation $(k,l)$ of the Fuzzy sphere. In other words, \eqref{e4.58} is \emph{exact} in the mass deformation parameter $\mu$ which follows from \eqref{e4.41}.

\paragraph{A summary.} In summary, we have described two parallel approaches to Krylov operator growth in the matrix model based on recursion relations. The traditional approach is the construction of the Lanczos coefficients $b_n$ that come with the recursion relations \eqref{e3.9} and \eqref{e3.13}. In an alternative approach, we introduce the Laplace transform \eqref{e3.66} and associated recurrence relations \eqref{e3.65} and \eqref{e3.85}. This second line of approach further allows us to construct the resolvent which leads to the spectral function and finally to the density of states \cite{coleman_2015}-\cite{agd_1975}. The spectral function leads to collective excitations that are characterized by complex pole structure near the resonance and thereby are characterized as overdamped or diffusive modes that have no propagating degrees of freedom. The power law nature of the pole and the complex part of the self energy is quite indicative of the fact that these collective excitations are pertinent to non-Fermi liquid type transports at low frequencies. Finally, by identifying and isolating the cubic pole structure at low frequencies, we present a method to regularize the spectral function that results in a finite density of states. 
\section{Krylov variance and Krylov entropy}
\label{sec5}
Before we conclude this paper, it is worth mentioning some more physical entities that play crucial role in the context of chaos in quantum many body dynamics. The first is the notion of Krylov variance \cite{Nandy:2024evd} which refers to fluctuations in the Lanczos coefficients $b_n$. This plays an alternate measure of how fast the information is \emph{scrambled} and becomes chaotic in a strongly correlated medium. We address above issues for matrix model in this section. 

The first step is to generalize the definition \eqref{e3.14} by introducing the notion of Krylov complexity of degree $k$, which is defined as \cite{Muck:2022xfc}
\begin{align}
\label{e5.1}
 \braket{n^k}=  \mathcal{K}^{(k)}_{\mathcal{O}}(t)= \sum_{n}n^k |\varphi_n(t)|^2.
\end{align}

The Krylov variance \cite{Caputa:2021ori} is defined following the definition in statistical mechanics 
\begin{align}
    \sigma^2_{n}=\braket{n^2}-\braket{n}^2=\mathcal{K}^{(2)}_{\mathcal{O}}(t)-(\mathcal{K}_{\mathcal{O}}(t))^2.
\end{align}

A straightforward computation reveals at large $\mu$ and small time $t \sim 0$
\begin{align}
  &\sigma^2_{n}(k)|_{t \sim 0}=\sum_n n^2 |\varphi_n(t)|^2=4\left[1+\frac{ (k(k-2) +257) (k+25)^2}{(k (k+20)+341)^2}\right] \mu ^2 t^2+\mathcal{O}(\mu^4 t^4).
\end{align}

In other words, the variance scales quadratically with time at early time scale controlled by the mass parameter $\mu$. We identify this as the pre-chaotic universal behavior in the matrix model, which like in the case of $K$-complexity \eqref{e3.43} does not yet know whether the system is integrable or chaotic. Thee behavior $\sigma^2_{n}\sim \mathcal{K}_{\mathcal{O}}(t)$ indicates that the operator $\mathcal{O}(t)$ behaves like a highly localized wave packet. It would be nice to extend the above analysis for late times where one would expect the breakdown of the quadratic nature near the scrambling time. Below, we list variance for different choice of $k$
\begin{equation}
\label{e5.4}
  \sigma^2_{n}(k)=
    \begin{cases}
      9.46 \mu^2 t^2 & \text{PFS}~ (k=1/4)\\
      9.28 \mu^2 t^2 & \text{IFS}~(k=1).
    \end{cases}       
\end{equation}

Finally, we introduce the notion of Krylov entropy \cite{Barbon:2019wsy}-\cite{Rabinovici:2020ryf} for BMN matrix model at large mass gap. Krylov entropy is a relatively new and powerful tool to probe quantum information scrambling. As time progresses, the initial operator $|\mathcal{O}_0)=|K_0)$ passes over through the Krylov basis elements $|K_n)$ with certain probabilities $p_n(t)=|\varphi_n(t)|^2$. The Krylov entropy is defined as the Shannon entropy of this probability distribution
\begin{align}
\label{e5.5}
    \mathcal{S}_{\mathcal{O}}(t)=-\sum_n |\varphi_n (t)|^2 \log |\varphi_n (t)|^2.
\end{align}

In recent years, the Krylov entropy has become a useful tool to diagnose chaos in quantum many body systems. For example, in many body chaotic systems, at early times, the Krylov entropy grows linearly with time, $ \mathcal{S}_{\mathcal{O}}(t) \sim \lambda t$, where the constant of proportionality is intimately related to the Lyapunov exponent of the system. 

In a typical quantum many body system, the Krylov entropy starts from zero at $t=0$, that is, $ \mathcal{S}_{\mathcal{O}}(t=0)=0 $ and increases for $t>0$, as the operator $\mathcal{O}(t)$ delocalizes over different Krylov basis $\{K_n\}$. This is indicative of the fact that the initial information has been scrambled over complex quantum many body dynamics and the Krylov entropy is a measure of how difficult it is to bring that information back. 

In order to evaluate \eqref{e5.5}, one has to evaluate the wavefunctions $\varphi_n(t)$ carefully. To begin with, as a first approximation, we take the wave function \eqref{e3.33} equal to unity, as we explore the dynamics close to $t\sim 0$. This stems from the fact that the initial operator $\mathcal{O}_0$ is highly localized and does not contribute to Krylov entropy. Therefore, at early time $t \sim 0$, the LO contribution to the Krylov entropy comes from \eqref{e3.32}, that is, $\varphi_1(t)=b_1 t+\mathcal{O}(t^2)$.

Notice that in the large-$\mu$ limit, the fundamental energy scale driving the Heisenberg evolution is set by the mass deformation $\mu$. To systematically extract the non-trivial real-time dynamics of the wavefunctions without encountering oscillatory artifacts, we perform a rescaled analytic continuation (Wick rotation) parameterised as
\begin{align}
\label{e5.6}
\mu \to \frac{\mu}{2\sqrt{k+25}}\,, \qquad t \to i t \sqrt{k+25}\,.
\end{align}

The above scaling \eqref{e5.6} can be motivated in a two fold way. First, the transformation in \eqref{e5.6} combines an analytic continuation to imaginary time $t \to i t \sqrt{k+25}$ with a frequency renormalization $\mu \to \frac{\mu}{2\sqrt{k+25}}$. In real time, the early-time operator evolution is dominated by fast, high-frequency phase oscillations governed by the heavy mass gap $\mu$, that is, the fundamental mass gap dominates the Heisenberg time evolution $\mathcal{O}(t) = e^{i \mathcal{H} t} \mathcal{O}_0 e^{-i \mathcal{H} t}$. Standard perturbation theory in real time $t$ produces rapid oscillatory terms $\sim e^{\pm i \mu t}$, obscuring the net growth of wavefunctions $\varphi_n(t)$. Performing the rescaled Wick rotation \eqref{e5.6} maps these high-frequency real-time phase oscillations into real exponential growth modes \eqref{e5.8}. This transformation acts as a mathematical filter to isolate the true, non-oscillatory Lyapunov-like rate $\lambda \propto \mu$ driven by the mass gap. In summary, the rescaled analytic continuation serves as a ``systematic filter'' that converts these phase oscillations into exponential modes \eqref{e5.8}. Second, we notice that the first Lanczos coefficient evaluated at $t \sim 0$ (the single commutator ``tree-level'' value) is local and $k$-independent, that is, $b_1 = 2\mu$ \eqref{e3.30}. However, as $n$ increases along the Krylov chain $\mathcal{O}_n = [\mathcal{H}, \mathcal{O}_{n-1}]$, operator growth probes the global geometry of the Fuzzy Sphere vacuum\footnote{The entity $b_1$ represents the very first Lanczos coefficient evaluated near $t \sim 0$. It measures the local, short-time rate of operator growth under a single Heisenberg commutator $\mathcal{O}_1 = [\mathcal{H}, \mathcal{O}_0]$. Because it is a local, UV-dominated quantity, it depends solely on the fundamental mass gap scale $\mu$ set by the mass deformation and is completely independent of the global geometry parameter $k$. On the other hand, the entity $\bar{b}_1$ represents the effective scale to which higher-order Lanczos coefficients $b_n$ asymptote as the index $n$ increases along the Krylov chain. As $n$ grows, higher-order nested commutators probe the non-local, IR dynamics and global geometry of the background Fuzzy Sphere vacuum. In other words, as time $t$ grows, quantum information spreads deeper into the Krylov chain. The dynamics at later times are governed by higher-order coefficients $b_n$ ($n \gg 1$). The sequence of coefficients $b_n$ asymptotes to an effective scale $\bar{b}_1$. Because late-time dynamics probe large $n$ where $b_n \sim \bar{b}_1$, the Krylov differential equations governing the amplitudes $\varphi_n(t)$ inherit this background-dependent $\sqrt{k}$-scaling. Rescaling the parameter $\mu$ and performing the rescaled Wick rotation absorbs this $\sqrt{k}$ growth factor, allowing us to isolate the true non-oscillatory rate $\lambda \propto \mu$ across the entire chain without early-time phase oscillations obscuring the entropy growth.}. The higher-order Lanczos coefficients $b_n$ do not stay fixed at $2\mu$; instead, they asymptote to the effective background-dependent scale $\bar{b}_1 \sim \frac{\mu}{4\sqrt{2}}\sqrt{k^2-8k+25}$. Clearly, under the rescaling of $\mu$ \eqref{e5.6}, the effective scaling scales $\sim \sqrt{k}$ in the limit $k\gg 1$, that is, $\bar{b}_{1, \text{eff}} \sim \frac{\mu \sqrt{k}}{8\sqrt{2}}$. The rescaled time parameter $t\sqrt{k+25}$ \eqref{e5.6} absorbs this $\sqrt{k}$-dependence in the Krylov differential equations\footnote{This can be seen as follows. In the limit $k \gg 1$, this asymptotic scale simplifies to $\bar{b}_{1, \text{eff}} \sim \frac{\mu \sqrt{k}}{8\sqrt{2}}$. Substituting this back into the continuum version of the Krylov differential equation yields $\partial_t \varphi_n(t) \sim \left( \frac{\mu \sqrt{k}}{8\sqrt{2}} \right) \big( \varphi_{n-1}(t) - \varphi_{n+1}(t) \big)$. To remove the explicit background geometry factor $\sqrt{k}$ from the differential operators, a rescaled time variable $\tau$ is defined via the analytic continuation $\tau = t \sqrt{k + 25}$. Using the chain rule, time derivatives with respect to the original physical time $t$ convert to derivatives with respect to $\tau$. Considering $k\gg 1$, this finally yields $\partial_\tau \varphi_n(\tau) \sim \left( \frac{\mu}{8\sqrt{2}} \right) \big( \varphi_{n-1}(\tau) - \varphi_{n+1}(\tau) \big)$.}, regularizing the early-time expansion and isolating the linear growth rate $S_{\mathcal{O}}(t) \propto \mu t$ \eqref{e5.9}.

Under this parameterisation \eqref{e5.6}, the differential equation governing the first non-trivial amplitude $\varphi_1(t)$ simplifies into an analytically tractable form, yielding
\begin{align}
    \varphi_1(t>0)=\frac{2}{i}\frac{1}{(k+25)}\Bigg [e^{\frac{i \mu t}{2}(k+25)} -1+\frac{i\mu t}{2}(k+25)\Bigg].
\end{align}

Finally, Wick rotating back to the real time axis $t \rightarrow -i t/\sqrt{k+25}$ and reinstateding the original mass parameter $\mu \rightarrow \infty$, one finds at an early time $t>0$ 
\begin{align}
\label{e5.8}
  \varphi_1(t>0)\approx -\frac{2ie^{\mu t} }{(k+25)}  + \cdots
\end{align}
which yields the early time growth of the Krylov entropy as
\begin{align}
\label{e5.9}
  \mathcal{S}_{\mathcal{O}}(t>0)= \frac{8\mu t}{(k+25)^2}+\mathcal{O}(\mu^2 t^2).
\end{align}

The linear growth in time $t$ of the Krylov entropy \eqref{e5.9} is an artifact of the chaotic nature of the matrix model at a large mass gap \cite{Asano:2015eha},\cite{Huh:2024ytz}. As time progresses, an initially localized operator \eqref{e3.3} delocalize over the Krylov basis elements. The linear slope \eqref{e5.9} quantifies the rate with which the initial bit of quantum information spreads over the Hilbert space of operators. Notice that the rate of growth is purely controlled by the mass parameter $\mu$ and is different for different choices of the Fuzzy sphere vacuum, that is $k=1/4$ or $k=1$. The growth rate being directly proportional to the mass parameter $\mu$ demonstrates that the heavy mass gap introduces a scale $\mu^{-1}$ that drives scrambling at early times. Since the early time coefficient increases linearly with $\mu$, this implies that the late time Lanczos coefficients $b_n$ and the associated scrambling rates are also directly scaled by $\mu$. 
\section{Summary and conclusions}
\label{sec6}
Let us conclude this paper by re-emphasizing the main results of the manuscript. The major outcome could be divided mainly into three parts.\\\\
\uline{Summary of results:}

$\bullet$ In the first part, we discuss the moments and the return amplitude for the BMN matrix model at large mass gap $\mu \gg 1$. We generalize these expressions for both spread complexity of states and the Krylov operator growth in the matrix model. We also discuss the general structure of the orthogonal polynomials for both spread complexity of states and Krylov operator growth in the matrix model at large mass gap.

$\bullet$ The second major outcome is the use of the notion of the Krylov operator growth to calculate the retarded Green's function that leads to the spectral function of the matrix model at large mass gap. At higher frequency, the spectral function dies out quickly, which is interpreted as an artifact of the higher scattering rates. The low frequency behavior of the spectral function reveals the presence of ``collective'' modes that lead to a richer structure to the quantum many body dynamics. These collective modes appear to be relatively stable and non-propagating, which results in diffusion in the system. The power law behavior of the pole appearing in the spectral function further indicates that these collective excitations could be thought of as non-Fermi liquid at low frequency. We further discuss the possible regularization of the spectral function and the road to finite density of states.

$\bullet$ Finally, we discuss the issue of ``scrambling'' of quantum information in matrix model and in particular at early times. We compute two important probes in this regard, namely, the Krylov variance and the Krylov entropy. The linear rate of growth of Krylov entropy further ensures the quantum chaotic nature of the matrix model at large mass gap.

Beyond the specific technical evaluation of Lanczos coefficients and spectral densities, our analysis provides broad conceptual takeaways for gauge/gravity duality, quantum chaos, and thermalization in mass-deformed gauge theories. Below, we outline these key points.

     \textbf{Dampening the operator growth:} The explicit suppression and modification of operator growth by the mass deformation parameter $\mu$ clarifies how mass gaps deform information scrambling compared to standard conformal CFTs or SYK-like holographic models. The mass deformation $\mu$ acts as a frequency gap that shifts energy scales and suppresses off-resonant matrix couplings, thereby dampening the growth rate of the Lanczos sequence $b_n$ and slowing down the spreading of operators across the Krylov chain compared to the un-deformed ($\mu=0$) regime or conformal holographic CFTs. In other words, the suppression of operator growth at large $\mu$ shows how deforming a holographic gauge theory away from conformal invariance slows down the microscopic thermalization process.
     
     \textbf{IR collective modes:} The resonance-induced IR divergences in the spectral density highlight the emergence of long-lived, non-propagating diffusion or relaxation modes in the large mass limit ($\mu \gg 1$). Characterized by purely imaginary poles ($\omega_\pm \approx \pm i \epsilon \sqrt{5}$) and an unsuppressed lifetime scaling as $\tau \propto 1/\omega^3$, these low-energy collective excitations are protected by phase-space scattering suppression. This offers a non-perturbative, operator-theoretic handle on non-Fermi liquid transport, slow thermalization regimes, and hydrodynamic relaxation in mass-deformed holographic gauge theories.

      \textbf{Bulk gravitational dynamics and scrambling confinement:} In the holographic dual, the mass parameter $\mu$ sets the flux/curvature scale of the 11D pp-wave background, acting as a strong gravitational harmonic potential. At large $\mu$, non-linear bulk gravitational interactions are strongly suppressed, confining matrix excitations around fixed classical vacua (fuzzy spheres). Consequently, quantum information remains trapped in small subsectors of the operator space, preventing the maximal exponential scrambling typical of CFTs or SYK models and providing a concrete holographic realization of slow thermalization in non-conformal quantum gravity.
    
    \textbf{Role of vacuum geometry:}  By computing the early-time behavior of Krylov entropy and Krylov variance, we explicitly observe a linear early-time growth, confirming the onset of quantum chaos even in the presence of a large mass gap. Remarkably, our explicit construction shows how the Lanczos coefficients and orthogonal polynomials explicitly depend on the fuzzy sphere representation ($k=1/4$ vs $k=1$). This provides a quantitative framework to study how quantum chaos and information spreading depend on the background vacuum geometry in matrix theories.\\\\
\uline{Future directions:}

Before we conclude, it is worth mentioning some future directions related to the analysis performed in this paper that are worth pursuing in the near future. Below, we list some of these natural and compelling questions that one could pursue. 

$\bullet$ It would be nice to explore many body physics in the intermediate and in the small mass deformation. In particular, to compute the corrections due to the \emph{finite} mass gap $\mu$ and study the properties of the retarded Green's function and the associated spectral function at low frequencies. It would be nice to explore the interpolation of the spectral function with gradually decreasing mass gap.

$\bullet$ Related to the above point, it would be nice to perform an explicit computation of the complex self energy beyond the large $\mu$ approximation and in particular to employ standard techniques like finite coupling many body perturbation theory or the maximum entropy method. It would be indeed nice to explore if and how interactions can shift the low energy non-Fermi liquid pole to a well defined quasi-particle description.

$\bullet$ Although the linear growth \eqref{e5.9} signifies the onset of chaos, the direct estimation of the Lyapunov exponent $\lambda_L$ is a non-trivial task. To extract $\lambda_L$ directly from Krylov operator growth, further information is needed regarding -(i) the asymptotic growth of the Lanczos coefficients $b_n$ for large $n$ and (ii) the exponential growth of the Krylov operator complexity at intermediate scrambling time $t \sim t_s$ rather than at early times $t \sim 0$. In this context, one can also explore the Krylov variance at higher order in time by computing higher moments $\braket{n^k}$ and investigate the relationship between the Krylov variance, operator dispersion and the non-Fermi liquid transport.

We leave these issues for future investigations.
\paragraph{Acknowledgements :}
 The author would like to thank Debashree Chowdhury, Simon F. Ross and Carlos Nunez for their valuable comments on the draft. The author acknowledges the Mathematical Research Impact Centric Support (MATRICS) grant (MTR/2023/000005) received from ANRF, India. \\ 


\begin{thebibliography}{99}

\bibitem{Parker:2018yvk}
D.~E.~Parker, X.~Cao, A.~Avdoshkin, T.~Scaffidi and E.~Altman,
``A Universal Operator Growth Hypothesis,''
Phys. Rev. X \textbf{9}, no.4, 041017 (2019)
doi:10.1103/PhysRevX.9.041017
[arXiv:1812.08657 [cond-mat.stat-mech]].

\bibitem{Balasubramanian:2022tpr}
V.~Balasubramanian, P.~Caputa, J.~M.~Magan and Q.~Wu,
``Quantum chaos and the complexity of Krylov of states,''
Phys. Rev. D \textbf{106}, no.4, 046007 (2022)
doi:10.1103/PhysRevD.106.046007
[arXiv:2202.06957 [hep-th]].

\bibitem{Balasubramanian:2022dnj}
V.~Balasubramanian, J.~M.~Magan and Q.~Wu,
``Tridiagonalizing random matrices,''
Phys. Rev. D \textbf{107}, no.12, 126001 (2023)
doi:10.1103/PhysRevD.107.126001
[arXiv:2208.08452 [hep-th]].

\bibitem{Dymarsky:2021bjq}
A.~Dymarsky and M.~Smolkin,
``Krylov complexity in conformal field theory,''
Phys. Rev. D \textbf{104}, no.8, L081702 (2021)
doi:10.1103/PhysRevD.104.L081702
[arXiv:2104.09514 [hep-th]].

\bibitem{Avdoshkin:2022xuw}
A.~Avdoshkin, A.~Dymarsky and M.~Smolkin,
``Krylov complexity in quantum field theory, and beyond,''
JHEP \textbf{06}, 066 (2024)
doi:10.1007/JHEP06(2024)066
[arXiv:2212.14429 [hep-th]].

\bibitem{Kundu:2023hbk}
A.~Kundu, V.~Malvimat and R.~Sinha,
``State dependence of Krylov complexity in 2d CFTs,''
JHEP \textbf{09}, 011 (2023)
doi:10.1007/JHEP09(2023)011
[arXiv:2303.03426 [hep-th]].

\bibitem{Alishahiha:2026fnu}
M.~Alishahiha and M.~J.~Vasli,
``Krylov distribution,''
Phys. Rev. D \textbf{113}, no.12, 126004 (2026)
doi:10.1103/bdlf-jjtw
[arXiv:2602.06150 [hep-th]].

\bibitem{Alishahiha:2022anw}
M.~Alishahiha and S.~Banerjee,
``A universal approach to Krylov state and operator complexities,''
SciPost Phys. \textbf{15}, no.3, 080 (2023)
doi:10.21468/SciPostPhys.15.3.080
[arXiv:2212.10583 [hep-th]].

\bibitem{Caputa:2025dep}
P.~Caputa and G.~Di Giulio,
``Local quenches from a Krylov perspective,''
JHEP \textbf{07}, 164 (2025)
doi:10.1007/JHEP07(2025)164
[arXiv:2502.19485 [hep-th]].

\bibitem{Caputa:2025mii}
P.~Caputa, G.~Di Giulio and T.~Q.~Loc,
``Growth of block-diagonal operators and symmetry-resolved Krylov complexity,''
Phys. Rev. Res. \textbf{7}, no.4, 4 (2025)
doi:10.1103/9v9v-54zv
[arXiv:2507.02033 [hep-th]].

\bibitem{Caputa:2025ozd}
P.~Caputa, G.~Di Giulio and T.~Q.~Loc,
``Symmetry-Resolved Krylov Complexity,''
[arXiv:2509.12992 [hep-th]].

\bibitem{Caputa:2024vrn}
P.~Caputa, H.~S.~Jeong, S.~Liu, J.~F.~Pedraza and L.~C.~Qu,
``Krylov complexity of density matrix operators,''
JHEP \textbf{05}, 337 (2024)
doi:10.1007/JHEP05(2024)337
[arXiv:2402.09522 [hep-th]].

\bibitem{Nandy:2024evd}
P.~Nandy, A.~S.~Matsoukas-Roubeas, P.~Mart{\'\i}nez-Azcona, A.~Dymarsky and A.~del Campo,
``Quantum dynamics in Krylov space: Methods and applications,''
Phys. Rept. \textbf{1125-1128}, 1-82 (2025)
doi:10.1016/j.physrep.2025.05.001
[arXiv:2405.09628 [quant-ph]].

\bibitem{Hashimoto:2023swv}
K.~Hashimoto, K.~Murata, N.~Tanahashi and R.~Watanabe,
``Krylov complexity and chaos in quantum mechanics,''
JHEP \textbf{11}, 040 (2023)
doi:10.1007/JHEP11(2023)040
[arXiv:2305.16669 [hep-th]].

\bibitem{Barbon:2019wsy}
J.~L.~F.~Barb{\'o}n, E.~Rabinovici, R.~Shir and R.~Sinha,
``On The Evolution Of Operator Complexity Beyond Scrambling,''
JHEP \textbf{10}, 264 (2019)
doi:10.1007/JHEP10(2019)264
[arXiv:1907.05393 [hep-th]].

\bibitem{Baggioli:2024wbz}
M.~Baggioli, K.~B.~Huh, H.~S.~Jeong, K.~Y.~Kim and J.~F.~Pedraza,
``Krylov complexity as an order parameter for quantum chaotic-integrable transitions,''
Phys. Rev. Res. \textbf{7}, no.2, 023028 (2025)
doi:10.1103/PhysRevResearch.7.023028
[arXiv:2407.17054 [hep-th]].

\bibitem{Alishahiha:2024vbf}
M.~Alishahiha, S.~Banerjee and M.~J.~Vasli,
``Krylov complexity as a probe for chaos,''
Eur. Phys. J. C \textbf{85}, no.7, 749 (2025)
doi:10.1140/epjc/s10052-025-14451-z
[arXiv:2408.10194 [hep-th]].

\bibitem{Bhattacharjee:2024yxj}
B.~Bhattacharjee and P.~Nandy,
``Krylov fractality and complexity in generic random matrix ensembles,''
Phys. Rev. B \textbf{111}, no.6, L060202 (2025)
doi:10.1103/PhysRevB.111.L060202
[arXiv:2407.07399 [quant-ph]].

\bibitem{Caputa:2024sux}
P.~Caputa, B.~Chen, R.~W.~McDonald, J.~Sim{\'o}n and B.~Strittmatter,
``Spread complexity rate as proper momentum,''
Phys. Rev. D \textbf{113}, no.4, L041901 (2026)
doi:10.1103/7zs8-9zpg
[arXiv:2410.23334 [hep-th]].

\bibitem{Heller:2024ldz}
M.~P.~Heller, J.~Papalini and T.~Schuhmann,
``Krylov complexity as holographic complexity beyond JT gravity,''
[arXiv:2412.17785 [hep-th]].

\bibitem{Heller:2025ddj}
M.~P.~Heller, F.~Ori, J.~Papalini, T.~Schuhmann and M.~T.~Wang,
``De Sitter holographic complexity from Krylov complexity in DSSYK,''
[arXiv:2510.13986 [hep-th]].

\bibitem{Fu:2025kkh}
Y.~Fu, H.~S.~Jeong, K.~Y.~Kim and J.~F.~Pedraza,
``Toward Krylov-based holography in double-scaled SYK,''
[arXiv:2510.22658 [hep-th]].

\bibitem{Rabinovici:2023yex}
E.~Rabinovici, A.~S{\'a}nchez-Garrido, R.~Shir and J.~Sonner,
``A bulk manifestation of Krylov complexity,''
JHEP \textbf{08}, 213 (2023)
doi:10.1007/JHEP08(2023)213
[arXiv:2305.04355 [hep-th]].

\bibitem{Erdmenger:2022lov}
J.~Erdmenger, A.~L.~Weigel, M.~Gerbershagen and M.~P.~Heller,
``From complexity geometry to holographic spacetime,''
Phys. Rev. D \textbf{108}, no.10, 106020 (2023)
doi:10.1103/PhysRevD.108.106020
[arXiv:2212.00043 [hep-th]].

\bibitem{Ambrosini:2024sre}
M.~Ambrosini, E.~Rabinovici, A.~S{\'a}nchez-Garrido, R.~Shir and J.~Sonner,
``Operator K-complexity in DSSYK: Krylov complexity equals bulk length,''
JHEP \textbf{08}, 059 (2025)
doi:10.1007/JHEP08(2025)059
[arXiv:2412.15318 [hep-th]].

\bibitem{Fatemiabhari:2025poq}
A.~Fatemiabhari, H.~Nastase, C.~Nunez and D.~Roychowdhury,
``Holographic Krylov complexity for conformal quiver gauge theories,''
Nucl. Phys. B \textbf{1025}, 117402 (2026)
doi:10.1016/j.nuclphysb.2026.117402
[arXiv:2512.14812 [hep-th]].

\bibitem{Roychowdhury:2026sgg}
D.~Roychowdhury,
``Krylov complexity for Lin-Maldacena geometries and their holographic duals,''
JHEP \textbf{05}, 197 (2026)
doi:10.1007/JHEP05(2026)197
[arXiv:2604.16977 [hep-th]].

\bibitem{Nastase:2026lhz}
H.~Nastase, C.~Nunez and D.~Roychowdhury,
``Holographic Krylov Complexity for Charged, Composite and Extended Probes,''
[arXiv:2604.07432 [hep-th]].

\bibitem{Fatemiabhari:2025usn}
A.~Fatemiabhari, H.~Nastase, C.~Nunez and D.~Roychowdhury,
``Holographic Krylov complexity in confining gauge theories,''
[arXiv:2511.22717 [hep-th]].

\bibitem{Fatemiabhari:2025cyy}
A.~Fatemiabhari, H.~Nastase and D.~Roychowdhury,
``Holographic Krylov complexity in N=4 SYM theory,''
Phys. Rev. D \textbf{113}, no.10, 106033 (2026)
doi:10.1103/6999-p31b
[arXiv:2511.19286 [hep-th]].

\bibitem{Baiguera:2025dkc}
S.~Baiguera, V.~Balasubramanian, P.~Caputa, S.~Chapman, J.~Haferkamp, M.~P.~Heller and N.~Y.~Halpern,
``Quantum complexity in gravity, quantum field theory, and quantum information science,''
[arXiv:2503.10753 [hep-th]].

\bibitem{Rabinovici:2025otw}
E.~Rabinovici, A.~S{\'a}nchez-Garrido, R.~Shir and J.~Sonner,
``Krylov Complexity,''
[arXiv:2507.06286 [hep-th]].

\bibitem{Berenstein:2002jq}
D.~E.~Berenstein, J.~M.~Maldacena and H.~S.~Nastase,
``Strings in flat space and pp waves from N=4 superYang-Mills,''
JHEP \textbf{04}, 013 (2002)
doi:10.1088/1126-6708/2002/04/013
[arXiv:hep-th/0202021 [hep-th]].

\bibitem{Asano:2015eha}
Y.~Asano, D.~Kawai and K.~Yoshida,
``Chaos in the BMN matrix model,''
JHEP \textbf{06}, 191 (2015)
doi:10.1007/JHEP06(2015)191
[arXiv:1503.04594 [hep-th]].

\bibitem{Amore:2024ihm}
P.~Amore, L.~A.~Pando Zayas, J.~F.~Pedraza, N.~Quiroz and C.~A.~Terrero-Escalante,
``Fuzzy spheres in stringy matrix models: quantifying chaos in a mixed phase space,''
JHEP \textbf{06}, 031 (2025)
doi:10.1007/JHEP06(2025)031
[arXiv:2407.07259 [hep-th]].

\bibitem{Roychowdhury:2026vzq}
D.~Roychowdhury,
``Krylov state complexity for BMN matrix model,''
[arXiv:2605.10786 [hep-th]].

\bibitem{Roychowdhury:2026igc}
D.~Roychowdhury,
``Krylov Complexity for Plane Wave Matrix Model,''
[arXiv:2605.26055 [hep-th]].

\bibitem{Huh:2024ytz}
K.~B.~Huh, H.~S.~Jeong, L.~A.~Pando Zayas and J.~F.~Pedraza,
``Krylov complexity in mixed phase space,''
Phys. Rev. D \textbf{111}, no.12, L121902 (2025)
doi:10.1103/gmy7-dn7l
[arXiv:2412.04963 [hep-th]].

\bibitem{Banks:1996vh}
T.~Banks, W.~Fischler, S.~H.~Shenker and L.~Susskind,
``M theory as a matrix model: A conjecture,''
Phys. Rev. D \textbf{55}, 5112-5128 (1997)
doi:10.1201/9781482268737-37
[arXiv:hep-th/9610043 [hep-th]].

\bibitem{Dasgupta:2002hx}
K.~Dasgupta, M.~M.~Sheikh-Jabbari and M.~Van Raamsdonk,
``Matrix perturbation theory for M theory on a PP wave,''
JHEP \textbf{05}, 056 (2002)
doi:10.1088/1126-6708/2002/05/056
[arXiv:hep-th/0205185 [hep-th]].

\bibitem{Ling:2006up}
H.~Ling, A.~R.~Mohazab, H.~H.~Shieh, G.~van Anders and M.~Van Raamsdonk,
``Little string theory from a double-scaled matrix model,''
JHEP \textbf{10}, 018 (2006)
doi:10.1088/1126-6708/2006/10/018
[arXiv:hep-th/0606014 [hep-th]].

\bibitem{Sugiyama:2002rs}
K.~Sugiyama and K.~Yoshida,
``Supermembrane on the PP wave background,''
Nucl. Phys. B \textbf{644}, 113-127 (2002)
doi:10.1016/S0550-3213(02)00794-0
[arXiv:hep-th/0206070 [hep-th]].

\bibitem{Lin:2005nh}
H.~Lin and J.~M.~Maldacena,
``Fivebranes from gauge theory,''
Phys. Rev. D \textbf{74}, 084014 (2006)
doi:10.1103/PhysRevD.74.084014
[arXiv:hep-th/0509235 [hep-th]].

\bibitem{Lin:2004kw}
H.~Lin,
``The Supergravity dual of the BMN matrix model,''
JHEP \textbf{12}, 001 (2004)
doi:10.1088/1126-6708/2004/12/001
[arXiv:hep-th/0407250 [hep-th]].

\bibitem{Asano:2014vba}
Y.~Asano, G.~Ishiki, T.~Okada and S.~Shimasaki,
``Emergent bubbling geometries in the plane wave matrix model,''
JHEP \textbf{05}, 075 (2014)
doi:10.1007/JHEP05(2014)075
[arXiv:1401.5079 [hep-th]].

\bibitem{Asano:2012zt}
Y.~Asano, G.~Ishiki, T.~Okada and S.~Shimasaki,
``Exact results for perturbative partition functions of theories with SU(2|4) symmetry,''
JHEP \textbf{02}, 148 (2013)
doi:10.1007/JHEP02(2013)148
[arXiv:1211.0364 [hep-th]].

\bibitem{Muck:2022xfc}
W.~M{\"u}ck and Y.~Yang,
``Krylov complexity and orthogonal polynomials,''
Nucl. Phys. B \textbf{984}, 115948 (2022)
doi:10.1016/j.nuclphysb.2022.115948
[arXiv:2205.12815 [hep-th]].

\bibitem{Muck:2024fpb}
W.~M{\"u}ck,
``Black holes and Marchenko-Pastur distribution,''
Phys. Rev. D \textbf{109}, no.12, 126001 (2024)
doi:10.1103/PhysRevD.109.126001
[arXiv:2403.05241 [hep-th]].

\bibitem{Caputa:2021ori}
P.~Caputa and S.~Datta,
``Operator growth in 2d CFT,''
JHEP \textbf{12}, 188 (2021)
[erratum: JHEP \textbf{09}, 113 (2022)]
doi:10.1007/JHEP12(2021)188

\bibitem{Barbon:2019wsy}
J.~L.~F.~Barb{\'o}n, E.~Rabinovici, R.~Shir and R.~Sinha,
``On The Evolution Of Operator Complexity Beyond Scrambling,''
JHEP \textbf{10}, 264 (2019)
doi:10.1007/JHEP10(2019)264
[arXiv:1907.05393 [hep-th]].

\bibitem{Rabinovici:2020ryf}
E.~Rabinovici, A.~S{\'a}nchez-Garrido, R.~Shir and J.~Sonner,
``Operator complexity: a journey to the edge of Krylov space,''
JHEP \textbf{06}, 062 (2021)
doi:10.1007/JHEP06(2021)062
[arXiv:2009.01862 [hep-th]].

\bibitem{coleman_2015}
Piers Coleman.
\newblock {\em Introduction to Many-Body Physics}.
\newblock Cambridge University Press, Cambridge, 2015.

\bibitem{agd_1975}
Alexei~A. Abrikosov, Lev~P. Gorkov, and Igor~E. Dzyaloshinski.
\newblock {\em Methods of Quantum Field Theory in Statistical Physics}.
\newblock Dover Publications, New York, 1975.
\newblock Translated by Richard A. Silverman.

\bibitem{bernevig_2013}
B.~Andrei Bernevig and Taylor~L. Hughes.
\newblock {\em Topological Insulators and Topological Superconductors}.
\newblock Princeton University Press, Princeton, NJ, 2013.

\bibitem{tinkham_1996}
Michael Tinkham.
\newblock {\em Introduction to Superconductivity}.
\newblock McGraw-Hill, New York, 2nd edition, 1996.

\bibitem{bruus_flensberg_2004}
Henrik Bruus and Karsten Flensberg.
\newblock {\em Many-Body Quantum Theory in Condensed Matter Physics: An
  Introduction}.
\newblock Oxford University Press, Oxford, 2004.
\end{thebibliography}
\end{document}